%% file: eurosys-2027.tex
\PassOptionsToPackage{table}{xcolor}
\documentclass[sigplan,10pt]{acmart}

\usepackage{epsfig,endnotes}
\usepackage{algorithm}
\usepackage[noend]{algpseudocode}
\usepackage{amsfonts}
\usepackage{amsmath}
\usepackage{hyperref}
\usepackage{url}
\usepackage{breakurl}
\usepackage{booktabs} 
\usepackage{float}
\usepackage{graphicx}
\usepackage{listings}
\usepackage{setspace}
\usepackage{xspace}
\usepackage{multirow}
\usepackage{hhline}
\usepackage{array}
\usepackage{tabularx}
\usepackage{semantic}
\usepackage{svg}
\usepackage{comment}
\usepackage[normalem]{ulem}
\usepackage{subcaption}
\usepackage{tcolorbox}
\tcbuselibrary{skins,breakable}

\usepackage{tikz}
\usetikzlibrary{tikzmark}

\usepackage{syntax}

\usepackage{enumitem}

\usepackage[font=normalsize]{caption}
\usepackage[export]{adjustbox}
\usepackage[skip=1ex,font=small]{caption}

\usepackage{pifont}
\usepackage{soul}
\usepackage{stmaryrd}

\usepackage{ragged2e}
\usepackage{circledsteps}

\AtBeginDocument{%
  }

\input{utils/user-defined-commands}

\acmBadgeR[https://www.acm.org/publications/policies/artifact-review-and-badging-current]{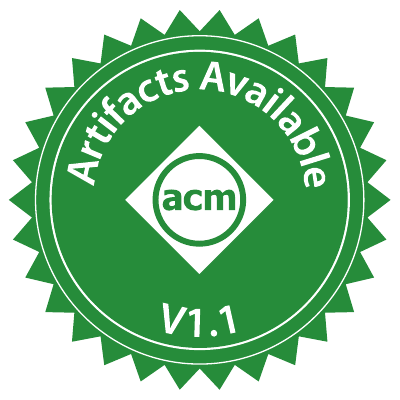}
\acmBadgeR[https://www.acm.org/publications/policies/artifact-review-and-badging-current]{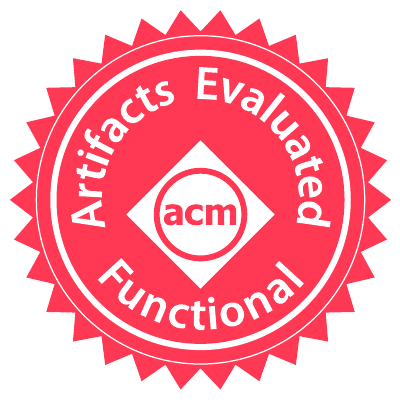}
\acmBadgeR[https://www.acm.org/publications/policies/artifact-review-and-badging-current]{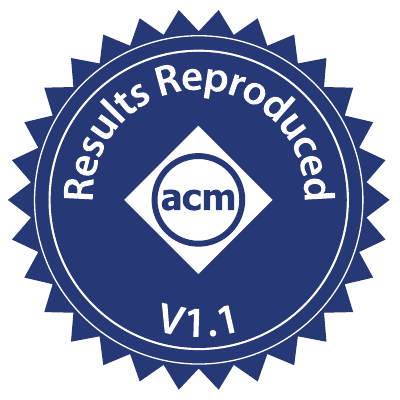}

\begin{document}
\acmSubmissionID{1508}
\title{\system: Instance-Specialized, Verifiable Systems Heuristics Through LLM-Driven Search}

\input{utils/authors}
\input{utils/abstract}
\input{utils/keywords}
\input{utils/copyright}

\settopmatter{printfolios=false}
\maketitle
\pagestyle{empty}

\input{sections/01-introduction}
\input{sections/02-background}
\input{sections/03-design-overview}
\input{sections/04-designing-search-space}
\input{sections/05-anvil}
\input{sections/06-thor-in-practice}
\input{sections/07-case-studies}

\input{sections/08-discussion}

\input{sections/09-related-work}
\input{utils/ack}

\bibliographystyle{ACM-Reference-Format}
\begingroup
\apptocmd{\thebibliography}{%
  \tolerance=200%
  \setlength{\parfillskip}{0pt plus 0.6\linewidth}%
}{}{}
\bibliography{refs}
\endgroup

\input{appendix/appendices}

\end{document}

%% file: utils/user-defined-commands.tex
\usepackage{xspace}
\usepackage{textcase}

\newcommand{\systemnameplain}{vulcan} 
\newcommand{\system}{\textsc{\MakeTitlecase{\systemnameplain}}\xspace} 
\newcommand{\libsystem} {\texttt{lib\MakeTitlecase{\systemnameplain}}\xspace} 

\newcommand{\dsl}{\textsc{Anvil}\xspace}

\newcommand{\Value}{\textsc{Value}\xspace}
\newcommand{\Rank}{\textsc{Rank}\xspace}

\newcommand{\ie}{\textit{i.e., }}
\newcommand{\eg}{\textit{e.g.,}\xspace}

\def\Snospace~{\S{}}

\newcommand{\mypara}[1]{\medskip\noindent{\bf {#1}}~}

\newcounter{packednmbr}

\newenvironment{packeditemize}{
\begin{list}{$\bullet$}{
\setlength{\itemsep}{0pt}
\addtolength{\labelwidth}{10pt}
\setlength{\leftmargin}{12pt}
\setlength{\listparindent}{\parindent}
\setlength{\parsep}{2pt}
\setlength{\topsep}{0pt}}}
{\end{list}
}

\newcommand{\circled}[1]{\CircledText[fill color=black, inner color=white]{#1}}

\usepackage{lstlinebgrd}
\tcbuselibrary{listings}
\lstnewenvironment{codebox}[1][]
  {\lstset{
    language=C++,
    basicstyle=\ttfamily\small,
    numbers=left,
    numberstyle=\tiny\color{black!60},
    numbersep=6pt,
    columns=fullflexible,
    keepspaces=true,
    showstringspaces=false,
    breaklines=true,
    frame=single,
    rulecolor=\color{black},
    aboveskip=0pt,
    belowskip=0pt,
    frame=single,
    morekeywords={i32,f32},
    commentstyle=\color{gray}\itshape,
    escapeinside={(*}{*)},
    #1
  }}
  {}

\newcommand{\highlightLine}[1]{\ifnum\value{lstnumber}=#1 \color{red!10}\fi}
\newcommand{\highlightLines}[1]{%
  \forcsvlist{\highlightLine}{#1}%
}

\newcommand{\highlightLineblue}[1]{\ifnum\value{lstnumber}=#1 \color{blue!10}\fi}

\newcommand{\constcolor}[1]{\textcolor{teal}{#1}}

\usepackage{fvextra}

\DefineVerbatimEnvironment{commandbox}{Verbatim}{
    breaklines=true,
    backgroundcolor=black!7,
    backgroundcolorpadding=5pt,
    fontsize=\small,
    formatcom=
}

%% file: utils/authors.tex
\author{Rohit Dwivedula}
\affiliation{%
  \institution{The University of Texas at Austin}
  \city{Austin}
  \state{TX}
  \country{USA}
}

\author{Divyanshu Saxena}
\affiliation{%
  \institution{The University of Texas at Austin}
  \city{Austin}
  \state{TX}
  \country{USA}
}

\author{Sujay Yadalam}
\affiliation{%
  \institution{The University of Texas at Austin}
  \city{Austin}
  \state{TX}
  \country{USA}
}

\author{Eric Hayden Campbell}
\affiliation{%
  \institution{The University of Texas at Austin}
  \city{Austin}
  \state{TX}
  \country{USA}
}

\author{Daehyeok Kim}
\affiliation{%
  \institution{The University of Texas at Austin}
  \city{Austin}
  \state{TX}
  \country{USA}
}

\author{Aditya Akella}
\affiliation{%
  \institution{The University of Texas at Austin}
  \city{Austin}
  \state{TX}
  \country{USA}
}

\renewcommand{\shortauthors}{Dwivedula et al.}

%% file: utils/abstract.tex
\begin{abstract}
Systems resource management tasks rely primarily on hand-designed heuristics. However, growing hardware heterogeneity and workload diversity require heuristics specialized to particular deployment instances, making manual design expensive and difficult to scale.
In this paper, we explore how to synthesize systems heuristics using LLMs. The main challenge is ensuring that generated heuristics execute safely, integrate correctly with the surrounding system, and still achieve strong performance.
We propose \system, a framework that identifies LLM-friendly interfaces that isolate core decision logic from the rest of the implementation.
With \system, LLM-generated code is restricted to simple stateless decision functions, while trusted runtime abstractions provide rich derived statistics for meaningful policy exploration without system-integration bugs.
To ensure execution safety, LLMs synthesize heuristics in a restricted language, \dsl, that guarantees important properties by construction.
We evaluate \system across three well-studied domains and demonstrate up to 4.9$\times$ higher savings for spot-VM scheduling, up to 2$\times$ lower miss ratios for cache eviction, and up to 14\% higher application performance for tiered-memory systems, while ensuring execution safety throughout.
\end{abstract}

%% file: utils/keywords.tex
\ccsdesc[500]{Information systems~Information storage systems}
\ccsdesc[500]{Software and its engineering~Operating systems}
\ccsdesc[300]{Computing methodologies~Randomized search}

\keywords{resource management heuristics, LLM-aided design, cache eviction, memory tiering, \mbox{cloud scheduling}}

%% file: utils/copyright.tex
\copyrightyear{2027}
\acmYear{2027}
\setcopyright{cc}
\setcctype{by}
\acmConference[EuroSys '27]{22nd European Conference on Computer Systems}{April 19--23, 2027}{Rabat, Morocco}
\acmBooktitle{22nd European Conference on Computer Systems (EuroSys '27), April 19--23, 2027, Rabat, Morocco}
\acmDOI{10.1145/3842654.3848582}
\acmISBN{979-8-4007-2971-3/2027/04}

%% file: sections/01-introduction.tex
\section{Introduction}

Systems resource management has long relied on manually designed heuristics.
This approach was sustainable when hardware was slower to evolve, deployment environments were more uniform, and a single carefully engineered policy could remain competitive across many settings.
Today, systems operate across rapidly evolving hardware platforms, increasingly heterogeneous memory and compute hierarchies, and workloads that vary across tenants, applications, and time~\cite{twitter-kv,tencent-dataset,alibaba-traces,azure-serverless-traces,dynamollm-hpca,tapas-asplos}.
As a result, the relevant unit of optimization is often an \emph{instance}: a particular combination of hardware, workload mix, deployment setting, and optimization objective.
The challenge is that designing and maintaining heuristics for each instance does not scale.

Neural models offered an early route to instance specialization~\cite{decima-sigcomm,orca,linnos,kleio-hpdc19,tcp-vivace-online-learning,darwin}.
A learned policy can, in principle, capture complex dependencies in system state and adapt to a target environment better than a fixed hand-written rule.
In practice, however, neural approaches remain difficult to deploy in core systems paths: their behavior is opaque, their training and serving pipelines add operational complexity, and integrating them with existing systems code is cumbersome~\cite{lake,guardrails-hotos}.
More broadly, systems heuristics do not exist in isolation.
They read state through existing interfaces, maintain auxiliary state over time, and rely on surrounding mechanisms to enact their decisions.
Any automated approach must therefore satisfy a substantially higher bar than predictive quality alone: it must integrate correctly with the surrounding system and be safe to execute.

Recent advances in large language models suggest a different route.
Unlike neural policies that must be embedded as external predictors, LLMs can synthesize \emph{code}, enabling instance-specialized heuristics to be expressed as executable logic compiled directly into the host implementation.
While enticing, this presents difficult challenges.
If the LLM has unrestricted control over heuristic generation, it must reason not only about policy quality but also about state management, memory safety, API semantics, synchronization, and mechanism interaction (\autoref{sec:why-llms-fail}).
Our experience, and that of recent work~\cite{policysmith,barbarians}, suggests that current LLMs can often discover performant heuristics under such freedom, but do not reliably produce code that satisfies the safety and integration requirements needed for deployment (\autoref{sec:unconstrained}).
Conversely, constraining the LLM too narrowly preserves safety but sharply limits the policy space it can explore (\autoref{sec:strawman-interface}).

This paper asks: \textit{how should resource-management code be organized so that instance-specialized heuristic synthesis is practical?}
Our answer is to refactor heuristics around a \emph{synthesis boundary} that isolates core decision logic from the rest of the implementation.
The key insight is that a small set of interfaces and runtime abstractions can make such synthesis both expressive and safe inside real systems.
Concretely, we show that the core decision logic of a broad class of systems heuristics can be expressed through two interfaces, \Rank and \Value, and that constructing rich statistics for performant policies can be delegated to reusable feature-store abstractions.
This organization permits the synthesis of verified policy code while keeping stateful logic and system-facing interaction in trusted, developer-managed code.

We realize this design in \system, a framework for synthesizing instance-specialized heuristics as safe executable code.
\system rests on two key ideas.
First, it decouples policy logic from state management.
Developers retain responsibility for high-level task decomposition, policy triggers, raw feature collection, and the mechanisms that enact decisions; LLM-synthesized artifacts are restricted to \textit{stateless decision functions} behind \Rank and \Value interfaces (\autoref{sec:design:interfaces}).
To enable meaningful policy exploration, \system exposes rich derived statistics through \libsystem, a library of \textit{listeners}: modular blocks that allow synthesized policies to use moving averages, percentiles, and other aggregates without implementing the data structures themselves (\autoref{sec:design:listeners}).
Second, \system synthesizes verified decision logic through \dsl, a restricted domain-specific language (DSL) for LLM-generated code (\autoref{sec:design:dsl}).
By excluding heap allocation, pointers, recursion, and unbounded loops, \dsl makes important execution safety properties such as memory safety, leak freedom, and termination \textit{provable by construction} for any well-formed program. Notably, the abstractions introduced by \system allow \dsl to remain restrictive while supporting powerful~heuristics.

This design occupies a different point in the design space from both prior LLM-based heuristic synthesis and prior ML-for-systems work.
Rather than asking the LLM to generate an entire heuristic implementation, \system narrows synthesis to a structured policy layer that remains expressive enough for meaningful heuristic discovery.
Rather than embedding an opaque learned model, \system produces explicit policy code that integrates with existing implementations through narrow interfaces and trusted runtime support.
We evaluate \system across multiple resource-management settings: deadline-aware spot-VM scheduling, cache eviction, and memory tiering.
Our findings show that \system can synthesize both general-purpose and (when required) instance-specialized heuristics that are competitive with, and in several cases outperform, strong human-designed baselines, while satisfying execution-safety properties by~construction.

In summary, we make the following contributions:
\begin{packeditemize}
    \item We identify a refactoring of systems heuristics around a synthesis boundary and design a framework, \system, that exposes only core decision logic to LLM-driven search while retaining system-facing structure in trusted code.
    \item We present \libsystem, which includes \Rank/\Value interfaces for decision logic and listener abstractions, which together provide a common substrate for expressive specialized systems heuristics.
    \item We introduce \dsl, a restricted policy language that yields execution-safety properties by construction, enabling practical deployment of LLM-generated heuristics.
    \item We evaluate \system for three well-studied domains: for spot VM scheduling, we find a \textit{general} heuristic that outperforms all baselines to yield up to 4.9$\times$ savings; for the well-studied cache eviction problem, we find \textit{specialized} heuristics that yield up to a 2$\times$ reduction in miss ratios compared to baselines for specific instances; for tiered-memory systems, we find page promotion heuristics that yield up to a 14\% improvement in application performance; all while satisfying important execution safety properties.
\end{packeditemize}

%% file: sections/02-background.tex
\section{The Pursuit of Instance Specialization}
\label{sec:bg}

Heuristics have been hand-crafted for various systems management tasks
because optimal actions are often intractable to compute online.
Action spaces for these tasks are large, decision intervals are fine-grained, and important system variables are often latent or partially observable.
Consequently, practitioners design heuristics for specific \emph{instances} of a problem, \ie particular combinations of workload, hardware, operating conditions, and optimization objectives such as performance, fairness, or utilization.
Seen this way, systems research is often a search for \emph{instance-specialized} heuristics rather than universally optimal ones.

\subsection{An Example of Specialized Heuristics: Caching}
\label{subsec:bg:instance-optimality}

We illustrate this search for specialized heuristics using the example of cache eviction policies.
Different heuristics have been proposed for specific workloads, objectives, or deployment scenarios~\cite{twoq, lhd, arc, sieve, halp, s3-fifo, cacheus, lecar}: some~\cite{arc,sieve} perform well for large cache sizes, while others~\cite{twoq,lhd} are more suited for smaller caches.
Workload characteristics also matter: scan-heavy workloads (mostly new objects) and churn-heavy workloads (mostly repeated objects) require different algorithms~\cite{cacheus}.
Heuristics have further been tailored for end-to-end \textit{objectives} such as tail latency~\cite{robinhood-osdi18} and fairness~\cite{robus-sigmod17}, and for system-level \textit{constraints} such as CPU overhead~\cite{halp}, lock-free design~\cite{s3-fifo}, and memory efficiency~\cite{tinyLFU}.

\begin{figure}[tb]
    \centering
    \includegraphics[width=1.0\linewidth]{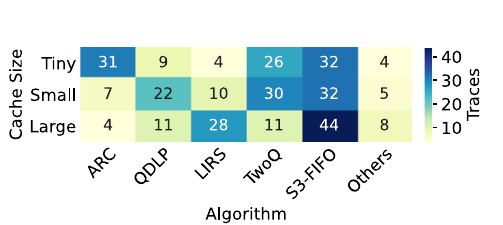}

    \caption{\textbf{Count of CloudPhysics traces where each heuristic performs best (max. object hit rate).} Tiny, small, and large imply cache sizes of 0.1\%, 1\%, and 10\% of trace footprint.}
    \Description{Heatmap showing, for each cache size, the number of CloudPhysics traces on which each cache-eviction algorithm achieves the highest object hit rate. Rows correspond to tiny, small, and large caches, equal to 0.1\%, 1\%, and 10\% of the trace footprint, respectively; columns correspond to ARC, QDLP, LIRS, TwoQ, S3-FIFO, and other algorithms. The winning algorithm varies substantially across traces and cache sizes: ARC performs best most often for tiny caches, while LIRS performs best most often for large caches, and no single eviction policy is best across all instances.}
    \label{fig:cloudphysics-motiv}
\end{figure}

To quantify this observation, we executed 17 caching algorithms using libCacheSim~\cite{libcachesim} on 106 block I/O traces from the CloudPhysics dataset~\cite{cloud-physics-dataset}, where each trace represents a distinct instance (corresponding to a different tenant).
\autoref{fig:cloudphysics-motiv} shows our findings: no single algorithm performs best on even half the instances, and the set of competitive algorithms shifts significantly with cache size, \eg ARC is better for tiny caches while LIRS is better for large caches.


These findings are not limited to cache eviction: policies across the systems stack exhibit the same instance dependence.
Congestion control algorithms are tailored for Internet~\cite{bbr,copa,orca,c2tcp} versus datacenter workloads~\cite{dctcp,swift,timely}, kernel queueing disciplines are tuned for varying performance and fairness objectives~\cite{qdisc-scrr,qdisc-stfq,qdisc-drr,qdisc-stratified-rr}, and cluster schedulers are fine-tuned for specific application classes such as data processing jobs~\cite{graphene,carbyne}, cloud VM workloads~\cite{lava-mlsys}, or deep learning training jobs~\cite{gandiva-osdi,gavel-osdi}.

However, as workloads grow increasingly diverse, performance objectives become multi-dimensional, and hardware heterogeneity evolves rapidly, manually specializing heuristics for each instance is no longer tractable.
This raises a natural question: \textit{can heuristic specialization be automated?}

\subsection{LLM-Based Synthesis: Requirements and Pitfalls}
\label{sec:why-llms-fail}

\begin{figure}[tb]
    \centering
    \includegraphics[width=0.8\linewidth]{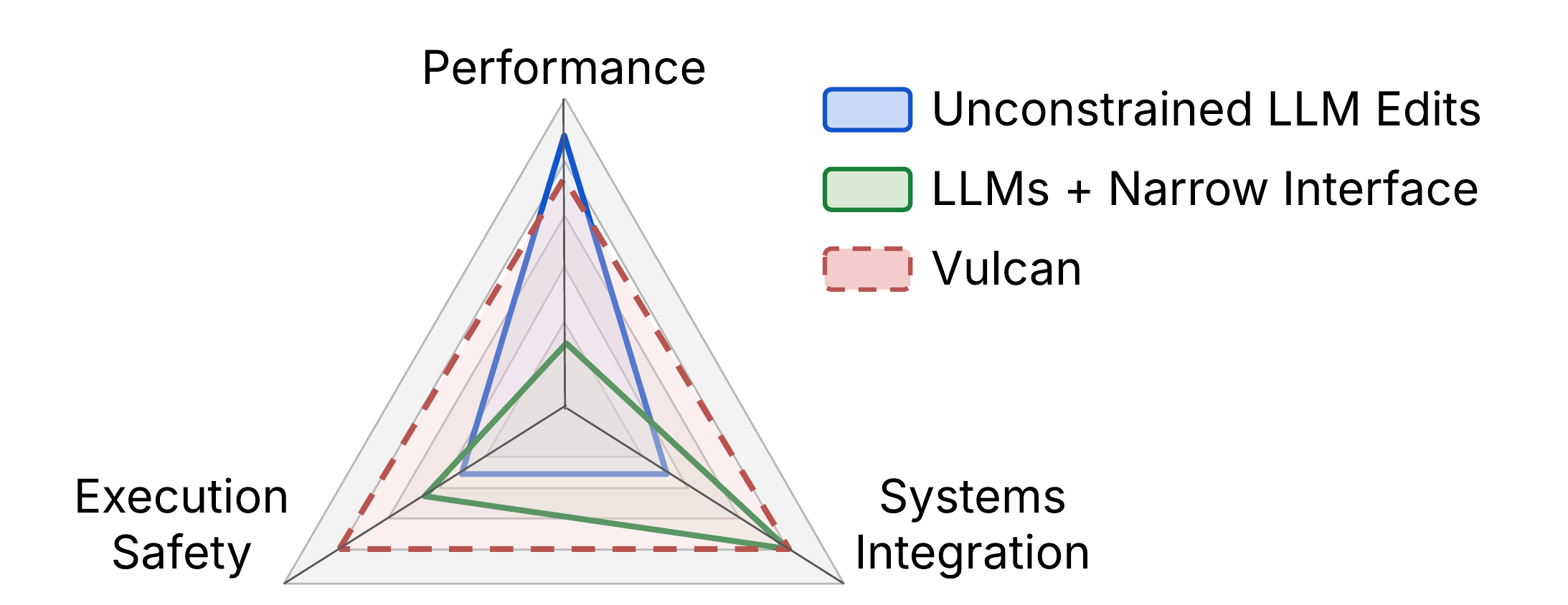}
    \caption{\textbf{Illustration of three design alternatives.} Unconstrained LLM edits result in execution safety and systems integration concerns (\autoref{sec:unconstrained}); overly constrained synthesis fails to achieve good performance (\autoref{sec:strawman-interface}). \system attempts to satisfy all three.}
    \Description{Radar chart comparing three approaches to LLM-based heuristic synthesis across performance, execution safety, and systems integration. Unconstrained LLM edits score highly on performance but poorly on execution safety and systems integration; an LLM with a narrow interface scores highly on execution safety and systems integration but poorly on performance; Vulcan is designed to achieve high performance while also preserving execution safety and systems integration.}
    \label{fig:triangle-of-deployability}
\end{figure}

Recent advances in LLMs and coding agents, particularly in generating executable code~\cite{reflexion-agent, tree-search-language-agent, executable-code-agents}, refining it using natural-language feedback~\cite{swe-agent}, and carrying out multi-step reasoning~\cite{chain-of-thought}, make automated heuristic specialization increasingly practical.
Indeed, several recent systems~\cite{alphaevolve,barbarians,glia,policysmith,robusta} have already demonstrated that LLM-based approaches can discover effective resource-management heuristics for specific problems.
While these advances are promising, building usable heuristics for practical systems imposes three key requirements.
\begin{packeditemize}
\item {\bf Performance:} They must meet or exceed the performance of existing manually designed heuristics.
\item {\bf Execution safety:} They must be \emph{proven} to be free from failures such as crashes, memory errors, leaks, or non-termination.
Automated synthesis at scale makes the conventional practice of manual review impractical, so safety verification must be automated as well.
\item {\bf System integration:}
Heuristics rarely run in isolation: they interact with other system components to read state and enact their decisions.
Thus, the generated code must respect the APIs and semantics of the surrounding system.
\end{packeditemize}

Naively applying LLMs may not satisfy all three properties; below, we explore the design space to identify an approach that does.
\autoref{fig:triangle-of-deployability} summarizes the alternatives, which we discuss in detail below.

\subsubsection{Unconstrained Search Space Approaches}
\label{sec:unconstrained}

At one extreme of the design space are existing proposals for LLM-driven heuristic synthesis~\cite{glia, barbarians, alphaevolve, policysmith}.
These approaches place few restrictions on what the LLM generates: starting from decision logic, the LLM agent is free to allocate and manage state and interact with system APIs as it sees fit.
This \emph{unconstrained search space} maximizes the LLM's opportunity to discover high-performing heuristics, but also exposes the full implementation surface to errors.

We illustrate this using cache eviction as a running example.
We compare two workflows: (i) an \emph{evolutionary loop}~\cite{funsearch,policysmith} that iteratively proposes, evaluates, and refines candidates; and (ii) an \emph{agentic} workflow with access to codebase inspection and other tools.
Both approaches used Claude Opus 4.5~\cite{claude-opus-4.5} to synthesize heuristics for libCacheSim~\cite{libcachesim}.
We evaluate the generated heuristics on ten traces from the CloudPhysics dataset~\cite{cloud-physics-dataset}.
Full details are provided in \autoref{app:why-llms-fail}; here, we discuss the key outcomes.

\mypara{Evolutionary loop.} We ran the search for fifty iterations and found that several synthesized heuristics had \textbf{system-integration violations}: several synthesized heuristics (including the best-performing one) under-reported per-object metadata size, claiming to use only 16 bytes while actually allocating up to 40 bytes.
libCacheSim uses reported metadata size to check whether the cache has space for a new object; under-reporting causes more objects to be admitted than the cache capacity allows, artificially inflating the hit rate.
We manually fixed these bugs and found that the resulting best heuristic achieved a 1.2\%--13.66\% improvement over FIFO, which is comparable to several manual baselines.

\mypara{Agentic workflow.} We allowed a Claude Code agent~\cite{claude-code} to make edits to the codebase until it reached a fixed cost budget, and repeated this ten times independently. Across these ten heuristics, four modified internal libCacheSim counters that are read-only -- another \textbf{system-integration violation}.
One heuristic also introduced a \textit{double-free bug}, an \textbf{execution safety violation};
incidentally, this specific bug was not triggered during testing, demonstrating that simple compile-and-test pipelines alone are insufficient.
The improvements for the best heuristic (out of ten) were 0.2--23\% over FIFO.

These results confirm that LLMs \textit{can} discover performant heuristics, but do not reliably produce code that satisfies safety and integration requirements.
Related work reports similar failures: ADRS~\cite{barbarians} noted use-after-erase and null dereference bugs, consistent with broader evidence that LLMs are less reliable on tasks requiring reasoning over complex codebases~\cite{bigcodebench, livecodebench, security-bugs-study-2026, llm-bugs-code-translation}.

\subsubsection{Overly Restricted Search Space Approaches}
\label{sec:strawman-interface}
The problems described in \autoref{sec:why-llms-fail} arise because the LLM has unrestricted access to the implementation surface.
The other extreme is to restrict LLM edits to narrow refinements of an existing heuristic~\cite{cc-tweak-bbr}, such that system integration and execution safety are preserved by construction.
This includes tasks such as adjusting threshold values, adding conditionals, or reweighting existing features.
However, shrinking the search space so aggressively forfeits the ability to discover substantially better policies.

To illustrate this, we revisit the evolutionary loop experiment for cache eviction but with a much narrower interface.
Inspired by HALP~\cite{halp}, we select eight eviction candidates using a \textit{base heuristic} and then restrict the LLM-written function to choosing the final candidate from within that set.
We provide the LLM with each candidate's access count, size, last-access timestamp, and insertion timestamp, plus percentile distributions of these statistics over all cached objects.
We experiment with both LRU and FIFO as base heuristics, running evolutionary search for 50 iterations each on the same traces as \autoref{sec:why-llms-fail}.
We refer to the resulting best heuristics as Evolved-LRU and Evolved-FIFO, respectively.

Under this interface, the LLM cannot interact directly with libCacheSim, eliminating system-integration violations and substantially narrowing the space of execution-safety failures.
Indeed, none of the synthesized heuristics exhibited execution-safety violations.
However, the same restriction limits the search space too much to produce meaningful gains: the best Evolved-LRU heuristic achieves a miss-ratio reduction (MRR) over FIFO by 0.05\%--7.76\%, nearly identical to LRU's 0.04\%--7.53\%.
Similarly, Evolved-FIFO improves over FIFO by only 0.01\%--0.19\%.
These gains are far smaller than those in \autoref{sec:why-llms-fail}.

\subsection{Towards a Better Division of Labor}
\label{sec:better-division-labour}
While LLM-synthesized heuristics hold promise for instance specialization, approaches at both extremes of the design space fail to meet all three requirements (\autoref{fig:triangle-of-deployability}), due to a poor division of labor between the LLM and the developer.
In the unconstrained setting (\autoref{sec:why-llms-fail}), the developer delegates everything to the LLM: decision logic, state management, and system interaction alike; in the over-restricted setting (\autoref{sec:strawman-interface}), the developer retains so much control that the LLM cannot meaningfully explore.
With \system, we identify abstractions that enable a better division of labor: supporting meaningful policy exploration while preserving execution safety and system integration by construction.

%% file: sections/03-design-overview.tex
\section{\system Overview}
\label{sec:design-overview}
\begin{figure}[tb]
    \centering
    \includegraphics[width=0.97\linewidth]{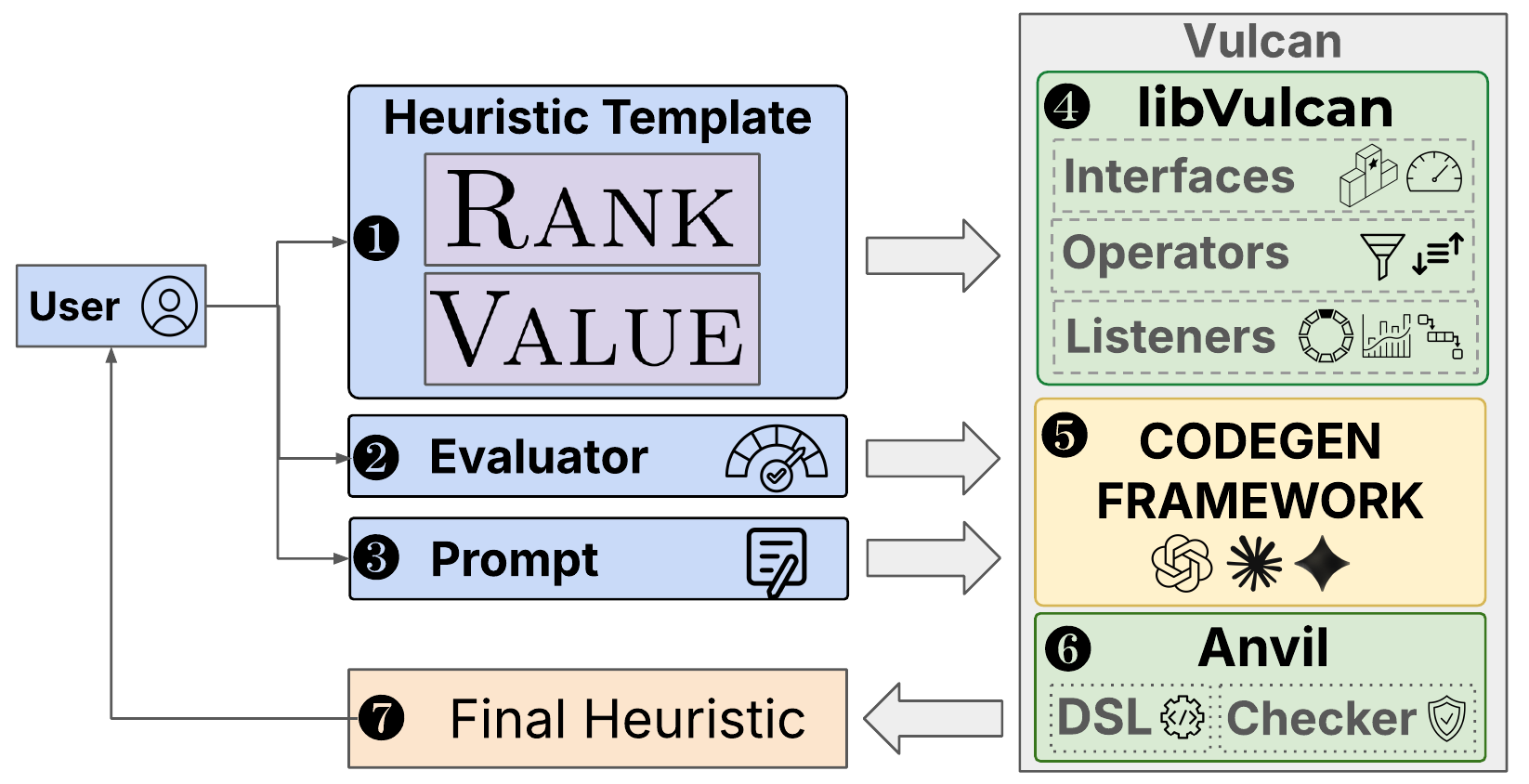}
    \caption{\textbf{Overview of \system.} Users provide task-specific inputs -- a template, a prompt, and an evaluator. \system uses a code-generation framework to produce a safe and performant heuristic.}
    \Description{Overview of the \system workflow. The user provides a heuristic template, evaluator, and natural-language prompt. The template uses \libsystem interfaces, operators, and listeners. A code-generation framework uses these inputs to synthesize candidate heuristics in the \dsl DSL, which checks their safety. Candidate heuristics are evaluated iteratively, and Vulcan returns the final heuristic to the user.}
    \label{fig:design-overview}
\end{figure}

We present \system, a framework for synthesizing safe, performant, and integration-compliant heuristics for systems resource-management tasks (see \autoref{fig:design-overview}).

\subsection{Key Ideas}
\label{sec:key-ideas}
\system rests on two key ideas that together address the three requirements from \autoref{sec:why-llms-fail}.

\mypara{Idea 1: Abstractions to decouple core decision logic and state management (\autoref{sec:design}).}
To explore meaningful heuristics, LLMs must derive useful statistics from raw features and use them to produce a decision.
\system decouples these two steps and restricts LLM edits to: (i) generating code for stateless decision interfaces, and (ii) selecting and querying an API for rich derived features.
We show that two stateless interfaces, \Rank (ranking candidates) and \Value (computing a scalar from system state), suffice to express the core decision logic of a broad class of resource-management heuristics (\autoref{sec:design:interfaces}).
A library of \textit{listeners} (\autoref{sec:design:listeners}) exposes useful statistics from raw features while handling the underlying state management.
All other aspects of system integration remain the responsibility of developers.

\mypara{Idea 2: Safety by construction via a restricted policy language (\autoref{sec:design:dsl}).}
As shown in \autoref{sec:why-llms-fail}, compile-and-evaluate pipelines alone are insufficient to ensure safety.
\system addresses this through \dsl, a restricted DSL in which LLMs express decision logic.
\dsl omits heap allocation, pointer arithmetic, recursion, and unbounded loops; as a result, memory safety, leak freedom, and termination are guaranteed by construction without the need for human review.

\subsection{Workflow: Using \system to Create Heuristics}
\label{sec:workflow}

Developers provide three inputs: a \emph{heuristic template} (\circled{1}), an \emph{evaluator} (\circled{2}), and a natural-language \textit{prompt} (\circled{3}) describing the resource-management task, the available raw features, and the evaluation methodology.

The heuristic template (based on Idea 1) exposes the core decision logic through \Rank or \Value interfaces and declares all raw features of interest.
Developers define this template using \libsystem (\circled{4}), \system's runtime library.
The template and prompt are passed to a \textit{code-generation framework} (\circled{5}), which synthesizes a heuristic in \dsl (\circled{6}) (based on Idea 2) by iteratively generating candidate heuristics, evaluating them on the user-provided evaluator, and using the feedback to improve.
LLMs may also attach listeners to raw features, deriving statistics from simple means to arbitrary percentiles.
The synthesized heuristic is passed to the evaluator, which scores it and returns feedback to the synthesis~loop.

Once the search terminates, \system returns the best-scoring candidate as the final heuristic (\circled{7}).
\system is agnostic to the code-generation approach; for our case studies (\autoref{sec:case-studies}), we use two approaches (ShinkaEvolve~\cite{shinkaevolve}, OpenEvolve~\cite{openevolve}).

%% file: sections/04-designing-search-space.tex
\section{\libsystem: Interfaces for Policy~Exploration}
\label{sec:design}
\system narrows LLM synthesis to the \emph{core decision logic} of a heuristic, delegating state management and system interaction to developer-owned code (\autoref{sec:design:division}).
This division leads to two concrete abstractions: stateless decision interfaces (\autoref{sec:design:interfaces}) and listener-based feature stores (\autoref{sec:design:listeners}).

\subsection{Division of Labor in \system}
\label{sec:design:division}

\begin{figure}[tb]
    \centering
    \includegraphics[width=\linewidth]{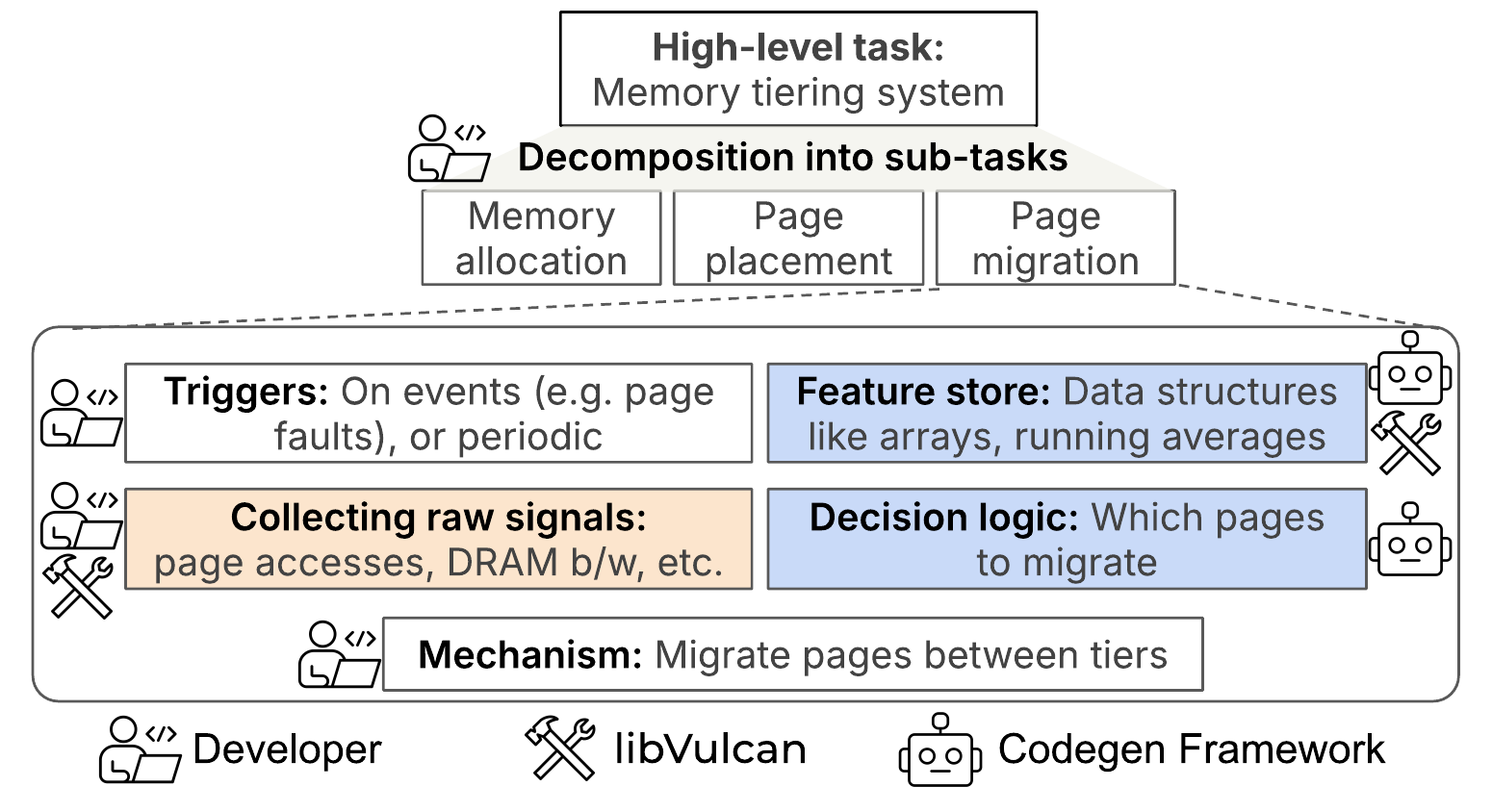}
    \caption{\textbf{Typical components of systems heuristics (using memory tiering as an example).} \system impacts the colored parts; icons show the division of responsibility among developers, \libsystem, and LLMs.}
    \Description{Diagram showing the division of responsibility in a memory-tiering heuristic. At the top, the high-level memory-tiering task is decomposed into memory allocation, page placement, and page migration; page migration is expanded into five components connected in sequence. The uncolored components -- task decomposition, triggers, raw-signal collection, and the migration mechanism -- are developer-managed. The colored components are those modified by \system: the feature store is managed by \libsystem, while the decision logic is synthesized by the code-generation framework. Icons beside each component indicate the responsible party: developer, \libsystem, or code-generation framework.}
    \label{fig:anatomy}
\end{figure}

To motivate the division of labor in \system, we examine the typical components of systems heuristics, illustrated in \autoref{fig:anatomy} using a memory-tiering system.

A high-level task is typically \textit{decomposed} into a collection of sub-policies; memory tiering, for example, uses separate policies for allocation~\cite{hemem}, placement~\cite{tpp2023,hemem}, and migration~\cite{arms} (\autoref{fig:anatomy}).
Within each policy, there are five typical components.
The \textit{trigger} determines when the policy runs, \eg page faults~\cite{hemem} or periodic scans~\cite{arms}.
Then, relevant statistics are \textit{collected}: for the page migration policy in \autoref{fig:anatomy}, these include page accesses and DRAM bandwidth.
Since processing raw signals at decision time can impose high overhead, a third component \textit{stores} them in data structures that support efficient computation of temporal or statistical aggregates; for instance, a windowed average of DRAM bandwidth rather than per-timestamp measurements.
The \textit{decision logic} consumes these derived statistics to produce a policy output.
Finally, \textit{mechanisms} enact the decision, \eg swapping pages between memory tiers.

With \system, decomposition, triggers, and mechanisms (uncolored in \autoref{fig:anatomy}) remain the developer's responsibility.
These components require deep knowledge of system semantics: in the memory-tiering example, decomposition depends on how allocators and page fault handlers interact; adding new triggers may require changes to kernel page fault handling; and implementing migration requires correctly managing page metadata, locks, and reference counts.
As \autoref{sec:why-llms-fail} showed, even state-of-the-art LLMs do not reliably handle this complexity.

\system automates the remaining components: feature storage and decision logic.
Developers expose raw signals through a heuristic template (\circled{1} in \autoref{fig:design-overview}), leaving open how those signals are aggregated and used in the decision.
The code-generation framework (\circled{5}) fills in this logic using \libsystem's APIs (\circled{4}) for rich derived statistics.

\mypara{Insight motivating \libsystem.}
\libsystem serves as the interface between developers, synthesized heuristics, and the system.
A key observation is that many safety and integration bugs arise when LLMs are responsible for managing low-level feature state: allocating buffers, maintaining data structures, and updating them correctly across events.
\libsystem eliminates this exposure by providing \textit{stateless interfaces} for decision logic (\autoref{sec:design:interfaces}) while encapsulating all state management in a \textit{listener library} (\autoref{sec:design:listeners}).

\input{tables/types-of-tasks}

\subsection{\Value and \Rank Interfaces}
\label{sec:design:interfaces}

The core decision logic of most resource-management heuristics takes one of two forms: it either \textit{ranks} a set of candidates and selects a subset (\eg evicting an object from cache), or it computes a scalar \textit{value} that drives a resource-management action (\eg computing a \texttt{cwnd} in congestion control).
\autoref{tab:types-of-policies} shows that this pattern recurs across a broad class of systems tasks, and \libsystem exposes two corresponding interfaces, \Rank and \Value, for LLM-generated decision logic.
When neither interface fits, it can be extended with a small number of additional functions.

\mypara{Interface: \Value.}
A \Value heuristic, $\textsf{Value}(X) \rightarrow Y$, is a function where $X$ encodes system state as a set of features and $Y$ is a scalar output corresponding to some control variable.
\autoref{tab:types-of-policies} lists examples of typical \Value tasks and their outputs.
For instance, a congestion controller maps features such as \texttt{rtt} and \texttt{loss\_rate} to a window size.
To use the \Value interface, developers define the function signature {\tt double value_fn (\systemnameplain::store\&)}, where {\tt store} contains the state $X$ and the return value is $Y$.

\input{tables/listeners}

\mypara{Interface: \Rank.}
A \Rank heuristic, $\textsf{Rank}(X, \mathcal{O}) \rightarrow \mathcal{R}$, is a function where $\mathcal{O} = \{o_i \mapsto f_i\}$ maps $N$ candidate objects to their features and $\mathcal{R} = [o_{i_1},o_{i_2},\dots]$ is an ordered list of all objects.
The mechanism then selects from this ordered list as needed.
Cache eviction, for example, ranks all objects in cache by their utility for retention (\autoref{tab:types-of-policies}) and evicts the lowest-ranked object.
As part of the template for \Rank tasks, developers define the function signature for a per-object scoring function: {\tt double score_fn (\systemnameplain::store\&, int64\_t obj\_id)}. \libsystem provides utility functions to sample candidates and sort them by this \texttt{score\_fn} in ascending or descending order as needed (\autoref{sec:impl:libvulcan}).

\subsection{Configurable Feature Stores Using \textit{Listeners}}
\label{sec:design:listeners}

Stateless decision functions cannot compute derived features such as moving averages, percentiles, or population-level statistics; yet smoothed averages~\cite{arms,tokencake,hyperbolic-caching,halp}, summary statistics~\cite{bbr,cubic}, and histograms~\cite{lhd} are essential to strong heuristics.
Delegating this state management to LLMs reintroduces the safety bugs from \autoref{sec:why-llms-fail}.

\system resolves this tension through \textit{listeners}: modular blocks that the LLM attaches to raw signals to create rich derived features.
Once attached, their query APIs can be called inside the decision logic, whether a value function (\Value) or a scoring function (\Rank).
\autoref{tab:libvulcan-listeners} summarizes the listeners provided by \libsystem.
Each listener tracks updates to its signal and maintains internal state for efficient queries.
For example, an {\tt Average} listener maintains running {\tt sum} and {\tt count} state and returns their ratio on query.
\libsystem also supports more complex aggregates: temporal aggregates (\eg {\tt RollingPercentile}) and population-level aggregates over all candidate objects (\eg {\tt PopMean}).
Temporal aggregates support both interfaces; population-level aggregates primarily support \Rank tasks.

Listeners need only be reviewed and verified for execution safety once, and a compact set of listeners can be reused across several systems tasks, as demonstrated by our case studies (\autoref{sec:case-studies}).
Listeners are also composable: multiple listeners can be attached to the same feature, and different features can use different listeners, forming a task-specific feature store without any LLM-written state management.

\input{code-snippets/scaffolding-example}

\subsection{An Example of Using \libsystem}
\label{sec:design:example}

\autoref{lst:e2e-example} demonstrates how the cache eviction task is set up using \libsystem. Lines~1--25 form the \textit{template}, which is one of the inputs to \system. This template is written by the user of \system (\ie the developer); LLMs in \system then search for instance-specialized heuristics within the confines of this template by deciding which listeners to add (lines~27--29) and defining the decision-making function (lines~31--35).

To create this template, the developer decides which raw signals the policy should use (lines~2--4), sets up instrumentation to collect these signals (within {\tt on\_insert} and {\tt on\_access}), and then formulates the problem using one or more of \system's interfaces (\Rank or \Value). In \autoref{lst:e2e-example}, cache eviction is formulated as a \Rank problem (lines~19--23), in which the lowest-ranked object is evicted; the developer can also choose between multiple implementations of \Rank provided by \libsystem (line~21, \autoref{sec:impl:libvulcan}).

\subsection{\libsystem Implementation}
\label{sec:impl:libvulcan}
\libsystem implements the \Rank and \Value interfaces as well as the listener APIs as a C++ library.
C/C++ systems can integrate \libsystem directly; Python-based systems use the same APIs via pybind11~\cite{pybind11} bindings. Our case studies exercise both integration paths: spot VM scheduling (\autoref{sec:spot-vm-scheduling}) uses Python bindings, while cache eviction (\autoref{sec:caching}) and memory tiering (\autoref{sec:tiering}) use the C++ library.

\libsystem provides multiple implementations of \Rank that the developer can choose between when setting up the template: {\tt OnDemandSort} (used in \autoref{sec:tiering}) scores and ranks candidates at decision time, while {\tt IncrementalSort} (used in \autoref{sec:caching}) precomputes ranks by maintaining candidates in a priority queue that is updated as features change, trading decision-time latency for incremental bookkeeping. These implementations also expose configurable parameters; \eg \texttt{OnDemandSort} can approximate ranks via a random sample of candidates rather than all of them (\autoref{lst:e2e-example}, line~21).

\libsystem totals $\sim$3.5K lines of C++ code: $\sim$1{,}300 for the core \Rank and \Value interfaces, $\sim$1{,}000 for listeners, and $\sim$250 for Python bindings.

%% file: tables/types-of-tasks.tex
\begin{table*}[t]
\footnotesize
\centering
\caption{\textbf{Examples of systems resource management tasks and the type of task (\Value or \Rank) they fall into.}
}
\begin{tabularx}{\linewidth}{@{}p{0.26\linewidth}Xp{0.26\linewidth}@{}}
\toprule
\textbf{Policy} & \textbf{Description} & \textbf{Type} \\
\midrule
Congestion control~\cite{cubic} & Decide the number of unacknowledged bytes in flight ({\tt cwnd}). & \Value: compute the {\tt cwnd}. \\
DVFS control~\cite{dvfs-sosp-2001} & Decide \texttt{cpu_freq} to balance performance \& power. & \Value: compute the frequency. \\
Cluster autoscaling~\cite{kubernetes-hpa} & Decide \texttt{n_replicas} to provision for a given service. & \Value: compute replicas per service.  \\
Hardware prefetching~\cite{hw-prefetching} & Choose an offset from the current access to prefetch data. & \Value: compute the offset value. \\
Cache eviction~\cite{lrb} & Select which cached object(s) to evict. & \Rank: all cached objects. \\
CPU scheduling~\cite{cfs-linux} & Select which thread to schedule next. & \Rank: all runnable threads. \\
Promotion in tiered memory~\cite{hemem,arms} & Select which pages to promote to a higher tier. & \Rank: all memory pages. \\
\bottomrule
\end{tabularx}
\label{tab:types-of-policies}
\end{table*}

%% file: tables/listeners.tex

\begin{table*}[t]
\centering
\footnotesize
\caption{\textbf{Listeners provided by \libsystem.} The table shows the state each listener maintains, with update/query/space complexities.
The first six rows show listeners for temporal aggregates, while the bottom two show listeners for aggregates across the object population.}

\begin{tabular*}{\linewidth}{@{\extracolsep{\fill}}lllccc@{}}
\toprule
\multicolumn{1}{@{}l}{\multirow{2}{*}{\textbf{Listener}}} & \multicolumn{1}{l}{\multirow{2}{*}{\textbf{Query API(s)}}} & \multicolumn{1}{l}{\multirow{2}{*}{\textbf{State and data structures}}} & \multicolumn{3}{c}{\textbf{Complexity}} \\
\cmidrule(l){4-6}
 & & & \textbf{Update} & \textbf{Query} & \textbf{Space} \\
\midrule
\texttt{Average()} & \texttt{get\_avg} & \texttt{sum}, \texttt{count} & $O(1)$ & $O(1)$ & $O(1)$ \\
\texttt{MinMax()} & \texttt{get\_min}, \texttt{get\_max} & \texttt{min}, \texttt{max} & $O(1)$ & $O(1)$ & $O(1)$ \\
\texttt{RollingWindow(N)} & \texttt{get\_kth}, \texttt{get\_avg} & ring buffer, sum & $O(1)$ & $O(1)$ & $O(N)$ \\
\texttt{EWMA($\{\alpha_0, \alpha_1, \ldots \alpha_M\}$)} & \texttt{get\_ewma$(\alpha)$} & current EWMA value for each $\alpha$ & $O(1)$ & $O(1)$ & $O(1)$ \\
\texttt{RollingPercentile(N)} & \texttt{get\_percentile} & ring buffer, ordered multiset~\cite{gnu-pbds} & $O(\log N)$ & $O(\log N)$ & $O(N)$ \\
\texttt{RollingCount(N)} & \texttt{get\_count}, \texttt{contains} & ring buffer, counts map & $O(1)$ & $O(1)$ & $O(N)$ \\
\midrule
\texttt{PopPercentile()} & \texttt{get\_pop\_percentile} & per-object latest, ordered multiset~\cite{gnu-pbds} & $O(\log M)$ & $O(\log M)$ & $O(M)$ \\
\texttt{PopMean()} & \texttt{get\_pop\_mean} & per-object latest, sum & $O(1)$ & $O(1)$ & $O(M)$ \\
\bottomrule
\end{tabular*}
\label{tab:libvulcan-listeners}
\end{table*}

%% file: code-snippets/scaffolding-example.tex
\begin{figure}[tb]
\begin{codebox}[basicstyle=\ttfamily\footnotesize,caption={\textit{Template} for cache eviction in \libsystem (lines~1--25). LLMs search for policies within this template; lines~27--35 (red) show a sample candidate policy synthesized during the search.},label={lst:e2e-example},linebackgroundcolor={\highlightLines{27, 28, 29, 30, 31, 32, 33, 34, 35}}]
void init():
  (*\systemnameplain*)::store store = {
    .obj_signals: [{"insert_time",int}, {"access_time",int}]
  };
  setup_listeners(store);

// LLM evolved functions
void setup_listeners((*\systemnameplain*)::store& fs);
double score_fn((*\systemnameplain*)::store& fs, int64 id);

void on_insert(ObjId o):
  store.add_obj(o);
  store.update("insert_time", o, now());

void on_access(ObjId o):
  store.update("access_time", o.id, now());

ObjId evict_obj():
  R = (*\systemnameplain*)::Rank(
    store.all_objs, score_fn,
    (*\systemnameplain*)::OnDemandSort, n_sample=32
  );
  victim = R.back();
  store.remove_obj(victim);
  return victim;

void setup_listeners((*\systemnameplain*)::store& fs):
  fs.add_listener("access_time",(*\systemnameplain*)::Latest());
  fs.add_listener("insert_time",(*\systemnameplain*)::PopPercentile());

double score_fn((*\systemnameplain*)::store& fs, int64 id):
  auto recency = fs.get_latest("access_time", id)-now()
  if fs.get_pop_percentile("insert_time", id) > 0.9:
      return 6.0 * recency
  else return 1.0 * recency
\end{codebox}
\end{figure}

%% file: sections/05-anvil.tex
\section{\dsl: DSL for LLM-Written Code}
\label{sec:design:dsl}
\input{tables/safety}
The abstractions in \autoref{sec:design} reduce the scope for execution-safety bugs but do not eliminate them.
\system closes this gap with \dsl: a domain-specific language in which every well-formed program satisfies the execution-safety properties in \autoref{tab:safety-properties} by construction.

\mypara{Programming model.}
\dsl is a restricted imperative language based on C++: it supports scalars, arithmetic, conditionals, bounded \texttt{for} loops, and built-in math functions (\eg \texttt{anvil::max}), while excluding pointers, arrays, containers, heap allocation, recursion, and unbounded loops.
These exclusions map directly to the execution safety properties (\autoref{tab:safety-properties}): excluding pointers and arrays prevents unsafe memory accesses (P1); excluding heap allocation and persistent state prevents memory leaks and state-management errors (P2); excluding recursion and unbounded loops prevents non-termination (P3).

\mypara{State management and trust model.}
Most useful systems heuristics rely on persistent state, such as histories or derived statistics, but pure \dsl does not contain the constructs needed to maintain such state. To let \dsl programs use stateful functionality implemented outside the language, \dsl supports \texttt{extern} declarations for C/C++ functions: the \dsl compiler sees the \texttt{extern}'s function signature (name, scalar arguments, and scalar return type), while its implementation resides in trusted external code. In \system, listener query APIs (\autoref{tab:libvulcan-listeners}) are exposed to synthesized heuristics through such \texttt{extern}s. This defines \dsl's trust model: a well-formed \dsl program satisfies P1--P3 provided every linked \texttt{extern} also satisfies P1--P3.

\mypara{Compiling and executing \dsl code.}
{In \system, code generators write heuristics in \dsl and use \libsystem functions to attach and query listeners (declared in \dsl as function prototypes).
The LLM-written \dsl code is then compiled in two stages. First, the frontend parses the generated heuristic against the \dsl grammar and rejects disallowed constructs. Next, the \dsl code is lowered to C and linked against C implementations of \libsystem functions (for attaching/querying listeners) that were simply declared as {\tt extern}s in the \dsl code. The final C implementation is then compiled with native toolchains (e.g., {\tt gcc}).
This lowering of \dsl code into C is made possible because of \dsl's C-like syntax: scalar expressions, conditionals, and bounded loops pass through unchanged, while \dsl built-ins (\eg \texttt{anvil::max}) resolve to equivalent C library calls.
Thus, executing \dsl code adds no interpreter or JIT overhead and makes integration with the rest of the codebase seamless. The only overhead is the compilation step that lowers \dsl code to C, which takes under one second per scoring function.}

\mypara{Implementation.}
\dsl is implemented in OCaml using Menhir~\cite{menhir} for parser generation, totaling approximately 3.5K lines of code: $\sim$1{,}000 for the parser and lexer, $\sim$1{,}500 for name resolution and class desugaring, and $\sim$1{,}000 for the AST and C code generation.

%% file: tables/safety.tex
\begin{table}[!t]
\caption{\textbf{Execution-safety properties guaranteed by \dsl.}}
\centering
\footnotesize
\begin{tabularx}{\linewidth}{@{}l>{\raggedright\arraybackslash}X@{}}
\toprule
\textbf{ID} & \textbf{Property} \\ \midrule
P1 & \textbf{Memory safety.} No out-of-bounds access, use-after-free, or~\mbox{double-frees}. \\
P2 & \textbf{Leak freedom.} No unbounded memory growth over time. \\
P3 & \textbf{Termination.} Must not run indefinitely; LLM scoring functions are guaranteed to eventually terminate. \\  
\bottomrule
\end{tabularx}
\label{tab:safety-properties}
\end{table}

%% file: sections/06-thor-in-practice.tex
\section{Using \system in Practice}
\label{sec:practice}

\mypara{Defining the target instance.}
A key requirement of \system-generated heuristics is high performance, which often requires instance specialization (\autoref{sec:bg}) across workloads, hardware configurations, operating regimes, or performance objectives.
When instances differ enough that no single heuristic performs well across them (\autoref{fig:cloudphysics-motiv}), policies should be specialized to a particular instance, such as a specific cluster, tenant, workload class, or arrival pattern (as we do for cache eviction in \autoref{sec:caching}).
In contrast, for understudied problems where good solutions are not yet well understood, even a single heuristic targeting average-case performance can yield substantial gains, as in our spot VM scheduling case study (\autoref{sec:spot-vm-scheduling}).

\mypara{Cost--fidelity tradeoff in evaluators.}
\system invokes the evaluator hundreds of times during search, so evaluation cost bounds the amount of exploration possible.
Evaluating heuristics on real systems can take hours per candidate; in practice, we use cheaper proxies during the search phase: trace-driven simulation (\autoref{sec:spot-vm-scheduling}, \autoref{sec:caching}) or execution over a short representative workload (\autoref{sec:tiering}).

\mypara{Natural-language prompt.}
As discussed in \autoref{sec:design-overview}, users must also provide a natural-language prompt. Besides describing the task at hand, this prompt must also include information about \libsystem APIs and may include evaluator details, workload or instance information, and instructions for heuristic design.

\mypara{LLM API budgets.}
In our experiments, we use Claude Sonnet 4.5 and Claude Opus 4.5 (80\%/20\% split) on AWS Bedrock to synthesize heuristics for \system and the baseline~\cite{openevolve, barbarians}.

%% file: sections/07-case-studies.tex
\section{Case Studies and Results}
\label{sec:case-studies}
\noindent We use \system to synthesize heuristics for three resource-management tasks:
\begin{packeditemize}
\item \textbf{Spot VM scheduling} (\autoref{sec:spot-vm-scheduling}): decide when a deadline-driven cloud job should use cheaper spot instances to minimize completion cost.
\item \textbf{Web cache eviction policies} (\autoref{sec:caching}): choose which cached objects to evict to maximize object hit rate ({\eg in CDNs}).
\item \textbf{Page promotion policy in tiered-memory systems} (\autoref{sec:tiering}): choose which pages to promote from slower memory tiers to faster tiers to maximize application-specific performance metrics.
\end{packeditemize}

\input{tables/results/case-study-descriptions}

\noindent The primary question we seek to answer through our evaluation is whether \system can synthesize heuristics that outperform strong, human-written baselines across these diverse tasks.
Our evaluation measures: (i) how \system compares to unconstrained LLM-based synthesis, (ii) how \system's components (interfaces, listeners, and the DSL) affect synthesized heuristics, and (iii) the cost of heuristic synthesis (\eg API charges and elapsed time) relative to the resulting benefits.
In subsequent sections, we will use these case studies and results to discuss the developer effort required to use \system (\autoref{sec:discussion:system-integration}) and the scope of its decision interfaces, including how they can be composed or extended to support additional tasks (\autoref{sec:discussion:interfaces}).

\input{sections/case-studies/spot-vm-scheduling}
\input{sections/case-studies/caching}
\input{sections/case-studies/memory-tiering}

%% file: tables/results/case-study-descriptions.tex
\begin{table*}[t]
\centering
\footnotesize
\caption{\textbf{Summary of use cases, baselines, evaluation metrics, and evaluators.}}
\begin{tabularx}{\linewidth}{@{}p{0.2\linewidth}p{0.32\linewidth}p{0.22\linewidth}X@{}}
\toprule
\textbf{Use case} & \textbf{Baselines} & \textbf{Metric} & \textbf{Evaluator} \\
\midrule
Spot VM scheduling (single) & Greedy, Uniform Progress~\cite{cant-be-late}, ADRS~\cite{barbarians} & Avg. USD saved & Simulator, 5\% sample \\
Spot VM scheduling (multi) & Round-robin uniform progress, ADRS~\cite{barbarians} & Avg. USD saved & Simulator, 5\% sample \\
Cache eviction & Nine state-of-the-art algorithms & Object miss-ratio reduction (MRR) & Simulator, 1M requests \\
Memory tiering & ARMS~\cite{arms} & Goodput, latency & Emulator, per-workload \\
\bottomrule
\end{tabularx}
\label{tab:use-cases}
\end{table*}

%% file: sections/case-studies/spot-vm-scheduling.tex
\subsection{Spot VM Scheduling}
\label{sec:spot-vm-scheduling}

Cloud providers offer two pricing tiers: on-demand instances are reliably available but expensive, while spot instances are up to 11$\times$ cheaper~\cite{cant-be-late} but can be preempted at any time.
Deadline-sensitive jobs therefore typically avoid spot instances and pay the on-demand premium for predictability.
Recent work~\cite{cant-be-late} proposed a scheduler that cuts this cost by opportunistically switching between spot and on-demand while still meeting deadlines.
{We use \system to synthesize scheduling heuristics for the single-region (\autoref{sec:spot-vm-scheduling-single}) and multi-region formulations (\autoref{sec:spot-vm-scheduling-multi}) of this problem and compare the resulting heuristics against unconstrained evolution~\cite{barbarians}.}

\subsubsection{Single-region scheduling.}
\label{sec:spot-vm-scheduling-single}

{In the single-region formulation~\cite{cant-be-late}}, the scheduler is initialized with per-job constants: the required VM type, execution time, deadline, and a \emph{changeover delay} cost incurred when the job is switched between on-demand and spot instances.
At each tick, the scheduler observes the current job status, progress so far, remaining work, and spot availability, then chooses one of three actions: run on spot, run on on-demand, or wait.

\underline{\textbf{Template}}.
{Since the per-tick decision in this problem is a single control variable encoding one of three actions, it naturally fits the \Value interface}, which maps observed state to a scalar output.
{While \Value supports arbitrary scalar outputs, we restrict its output domain here to three values corresponding to the actions above.}
The per-tick features and per-job constants are passed as inputs to the \Value function, and the LLM may attach listeners to the per-tick features.

\underline{\textbf{Manually designed heuristics}}. We use two baselines from Wu et al.~\cite{cant-be-late}.
The first, a \textit{greedy heuristic}, uses spot VMs opportunistically, waits when they are unavailable, and switches to on-demand only when further waiting risks missing the deadline.
The second is \textit{Uniform Progress} (UP), the state-of-the-art heuristic for this problem. Uniform Progress keeps the job on a linear progress trajectory from start to finish: it exploits spot VMs when they are available, but if needed, switches back to on-demand to catch up with the linear pace.

\underline{\textbf{Evaluator}}. We use the simulator and traces from Wu et al.~\cite{cant-be-late}: spot-availability data for four AWS instance types (1- and 8-GPU K80/V100) paired with job configurations to yield 21{,}600 evaluation cases.
During synthesis, candidates are scored by running a fixed 5\% sample of all evaluation cases and computing the \textit{average USD savings} over the greedy baseline; all reported results use the full 21{,}600-case set.

\underline{\textbf{Results}}.
We run OpenEvolve~\cite{openevolve} for 100 iterations and compare the resulting \system scheduler against two baselines: (i) UP, the manual baseline above, and (ii) ADRS~\cite{barbarians}, which runs OpenEvolve with the same budget but allows unconstrained edits to the scheduler policy.

\begin{figure}[tb]
    \centering
        \includegraphics[width=0.99\linewidth]{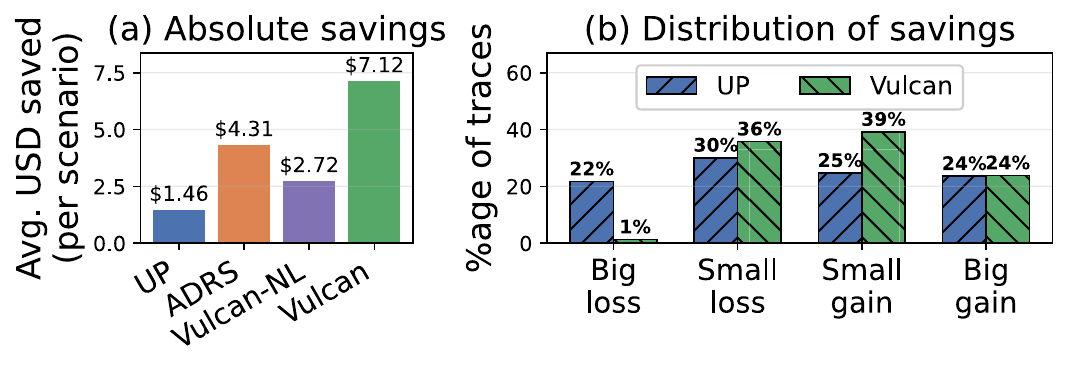}
    \caption{\textbf{Comparison of schedulers for the single-region spot VM setting}. (a) Mean savings; higher is better. (b) Big losses/gains exceed \$10; small losses/gains are below \$10.}
    \Description{Two panels compare schedulers for the single-region spot-VM setting. The first reports average USD savings per scenario relative to the greedy baseline: \$1.46 for UP, \$4.31 for ADRS, \$2.72 for Vulcan-NL, and \$7.12 for Vulcan. The second reports the distribution of savings for UP and Vulcan across four categories. UP has 22\% big losses, 30\% small losses, 25\% small gains, and 24\% big gains; Vulcan has 1\% big losses, 36\% small losses, 39\% small gains, and 24\% big gains. Vulcan therefore achieves substantially higher average savings while nearly eliminating big losses relative to UP.}
    \label{fig:spot-single-region-results}
\end{figure}

\noindent \textbf{Main result}. \autoref{fig:spot-single-region-results}a shows the mean savings of each scheduler over the greedy baseline. Our synthesized scheduler achieves nearly 4.9$\times$ the savings of UP and 1.7$\times$ those of ADRS.

\noindent \textbf{How \system outperforms UP.} The per-trace distribution (\autoref{fig:spot-single-region-results}b) reveals how \system outperforms UP: UP performs worse than the greedy heuristic (\ie costs more money) on 52\% of the evaluation cases.
This pattern stems from UP's core design assumption: it treats spot availability as \emph{worst-case unreliable}.
UP switches to on-demand whenever the job falls behind its expected progress schedule, \emph{regardless of spot availability history}.
In contrast, the \system heuristic (\autoref{lst:cbl-single}) attaches \texttt{RollingWindow} and \texttt{Average} listeners to track recent spot availability (\texttt{has\_spot}), classifying each tick into a \textit{scarce} or \textit{abundant} regime and adapting accordingly: committing to on-demand sooner when spot looks unreliable, but staying patient and returning opportunistically when it does not.

\input{code-snippets/spot-heuristics}

\begin{figure}[tb]
  \centering
  \includegraphics[width=0.9\linewidth]{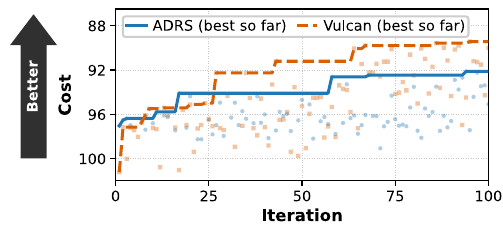}
  \caption{\system vs. unconstrained search (ADRS). Lines: best-performing-heuristic so far; dots: individual candidates.}
  \Description{Plot comparing Vulcan and unconstrained ADRS over 100 synthesis iterations. The horizontal axis is iteration number and the vertical axis is heuristic cost, where lower is better. Faint points show the costs of individual candidates and step lines show the best cost found so far. The two methods make similar progress during roughly the first 20 iterations, reaching best costs near 95 on the Y axis. Vulcan then improves more quickly, dropping to about 92 by iteration 25 and about 90 by iteration 60, while ADRS remains around 94 until roughly iteration 60 and finishes near 92. Vulcan therefore reaches its best solutions earlier and ends with a lower best cost than ADRS.}
  \label{fig:thor-vs-adrs-spot}
\end{figure}

\noindent \textbf{Unconstrained search.}
\autoref{fig:thor-vs-adrs-spot} compares \system to ADRS under an equal budget. Despite limiting the LLM to a \Value function written in Anvil (\autoref{sec:design:dsl}), \system loses no performance to unconstrained search---in fact, it reaches better heuristics faster. A possible explanation is the smaller, structured search space: while ADRS must explore the entire heuristic implementation, \system explores only the decision logic, enabling targeted policy changes.

\noindent \textbf{Impact of removing listeners.}
Without listeners (\system-NL), the heuristic is limited to the latest value of each feature.
\system-NL still outperforms UP, but full \system achieves roughly 2.6$\times$ the savings of \system-NL. {Without listeners, the scheduler cannot use recent spot-availability history to distinguish scarce and abundant regimes, preventing the adaptive behavior used by the full \system heuristic.}

\noindent \textbf{Cost of synthesis.} Synthesizing the heuristic (\autoref{lst:cbl-single}) with Vulcan incurred \$11 in LLM costs and took 2.5 hours (\autoref{tab:synth-cost}). Across all 21{,}600 evaluation scenarios (\ie simulated jobs), the resulting scheduler saves an average of \$7.12 per job relative to the greedy baseline, or \$2.81 more than the best ADRS solution. At this average improvement over ADRS, using the synthesized heuristic for just four jobs would recover the LLM synthesis cost on average.

\begin{table}[tb]
\centering
\footnotesize
\caption{\textbf{Cost of heuristic synthesis with \system.}}
\label{tab:synth-cost}
\begin{tabular*}{\linewidth}{@{\extracolsep{\fill}}lrrrr@{}}
\toprule
\multicolumn{1}{@{}l}{\multirow{2}{*}{\textbf{Use case}}} & \multicolumn{2}{c}{\textbf{Tokens (millions)}} & \multicolumn{1}{c}{\multirow{2}{*}{\textbf{API cost}}} & \multicolumn{1}{c}{\multirow{2}{*}{\textbf{Time}}} \\
\cmidrule(lr){2-3}
 & \multicolumn{1}{c}{\textbf{Input}} & \multicolumn{1}{c}{\textbf{Output}} & & \\
\midrule
Spot VM (single)              & 2.2      & 0.20       & \$11     & 2.5\,h \\
Spot VM (multi)               & 1.0      & 0.12       & \$5      & 1.2\,h \\
Cache eviction{$^\dagger$} & 0.7--2.0 & 0.15--0.50 & \$5--15  & 2.4--4.4\,h \\
Memory tiering{$^\dagger$} & 1.6--2.2 & 0.23--0.36 & \$10--14  & 1.9--3.0\,h \\
\bottomrule
\end{tabular*}
\par
{\footnotesize $\dagger$\, Minimum--maximum range across instances.}
\end{table}

\subsubsection{Multi-Region Scheduling.}
\label{sec:spot-vm-scheduling-multi}
In the multi-region formulation, the scheduler is allowed to choose which region to run the job in, with each region having different pricing and spot availability. {At each tick, in addition to choosing to schedule a job or wait, the scheduler can also choose to switch to a different region.} Since the scheduler can observe spot availability only for the \textit{current region}, a good heuristic would have to balance exploration and exploitation when choosing regions.

\underline{\textbf{Template}}.
Unlike the single-region case study, {this task does not directly fit the semantics of either \Rank or \Value: the scheduler output at each tick is a 2-tuple of (action, region).} We therefore decompose decision-making into two parts.
First, a \Value function, invoked once per scheduler tick, chooses the action: use spot instances in the current region, use on-demand instances in the current region, migrate to another region{, or wait}.
Second, if a migration action is selected, a \Rank function selects which region to migrate to.
{Although this region-choice \textit{could} also be expressed as a \Value task (with a fixed output domain of region identifiers), destination selection naturally involves \textit{comparing} regions based on their respective observation histories, making \Rank's semantics a more natural fit.}

\underline{\textbf{Evaluator}}.
We use the multi-region simulator and benchmark from ADRS~\cite{barbarians}, with 600 evaluation cases with AWS spot availability and prices for V100 GPU instances across nine U.S. regions/zones.
During evolution, the evaluator runs each candidate heuristic on a fixed 5\% sampled subset of all evaluation cases and returns the average cost per trace; after synthesis, the best policy is evaluated on the full benchmark.

\begin{figure}[tb]
    \centering
    \includegraphics[width=0.97\linewidth]{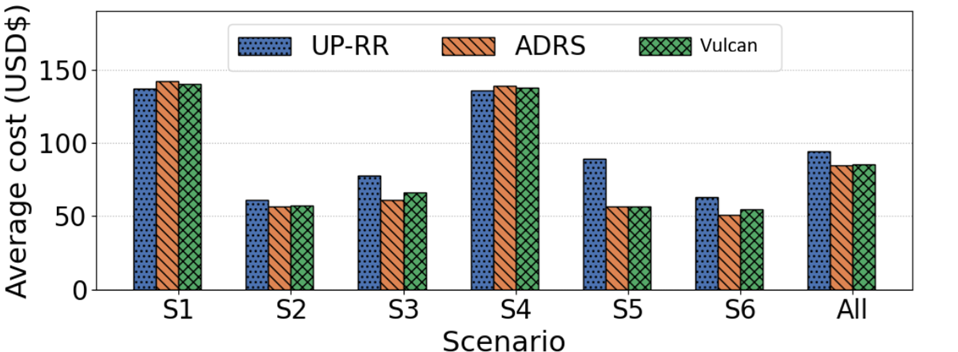}
    \caption{Average workload cost per scenario (lower is better).}
    \Description{Grouped bar chart comparing average workload cost for UP-RR, ADRS, and Vulcan across six multi-region scheduling scenarios (S1--S6), plus the average across all scenarios; lower cost is better. In S1 and S4, all three methods are close, at roughly \$140 per workload. In S2, Vulcan is lowest at about \$50, compared with roughly \$55 for ADRS and \$60 for UP-RR. In S3, ADRS is lowest at about \$60, followed by Vulcan at about \$68 and UP-RR at about \$80. In S5 and S6, Vulcan and ADRS are both near \$50--55, substantially below UP-RR at roughly \$90 and \$60, respectively. Averaged across all scenarios, Vulcan and ADRS are both around \$85, while UP-RR is around \$95.}
    \label{fig:spot-multi-region-results}
\end{figure}

\underline{\textbf{Results}}.
We run OpenEvolve for 100 iterations and compare the result against: (i) UP-RR~\cite{barbarians}, a round-robin multi-region extension of UP, and (ii) an unconstrained-search baseline (ADRS).
We report results for six scenarios~\cite{barbarians} covering different combinations of availability zones and geographic regions, each with 100 evaluation cases.
As \autoref{fig:spot-multi-region-results} shows, \system outperforms UP-RR in four of six scenarios, demonstrating how performant heuristics can be found even within constrained interfaces that guarantee execution safety. As in the single-region case study, synthesizing this multi-region heuristic required only 1.2 hours on a single node and \$5 in LLM API costs (\autoref{tab:synth-cost}).

Over all six scenarios, \system achieves only $\sim$7\% smaller savings than unconstrained search, demonstrating that its constrained interfaces can express performant heuristics while guaranteeing execution safety.


%% file: code-snippets/spot-heuristics.tex
\begin{figure}[tb]
\begin{codebox}[basicstyle=\ttfamily\footnotesize,caption={Regime detection in \system's scheduler.},label={lst:cbl-single}]
double alpha = fs.get_avg(has_spot);
// longest consecutive spot availability streak (L_max)
int64_t L_max = 0, L_curr = 0;
for (int i = 0; i < 18; ++i):
  if (fs.get_kth(has_spot, i) == 1):
    L_curr++; L_max = max(L_max, L_curr);
  else L_curr = 0;
// regime detection: scarce when alpha low or small streaks
bool scarce = (alpha < 0.35)||(L_max<od_ticks && alpha<0.5);
\end{codebox}
\end{figure}

%% file: sections/case-studies/caching.tex
\subsection{Cache Eviction}
\label{sec:caching}

\begin{table}[tb]
\caption{\textbf{Raw signals for cache-eviction scoring.}}
\centering
\footnotesize
\begin{tabularx}{\linewidth}{@{}lX@{}}
\toprule
\textbf{Feature type} & \textbf{Attributes} \\
\midrule
Per-object~($f_i$) & Access count, last access timestamp, insertion timestamp, size (bytes) \\
Global ($X$) & Current time, recently evicted object IDs (ghost list). \\
\bottomrule
\end{tabularx}
\label{tab:cache-features}
\end{table}

\begin{figure*}[t]
    \centering
    \begin{subfigure}[t]{0.49\textwidth}
        \centering
        \includegraphics[width=\linewidth]{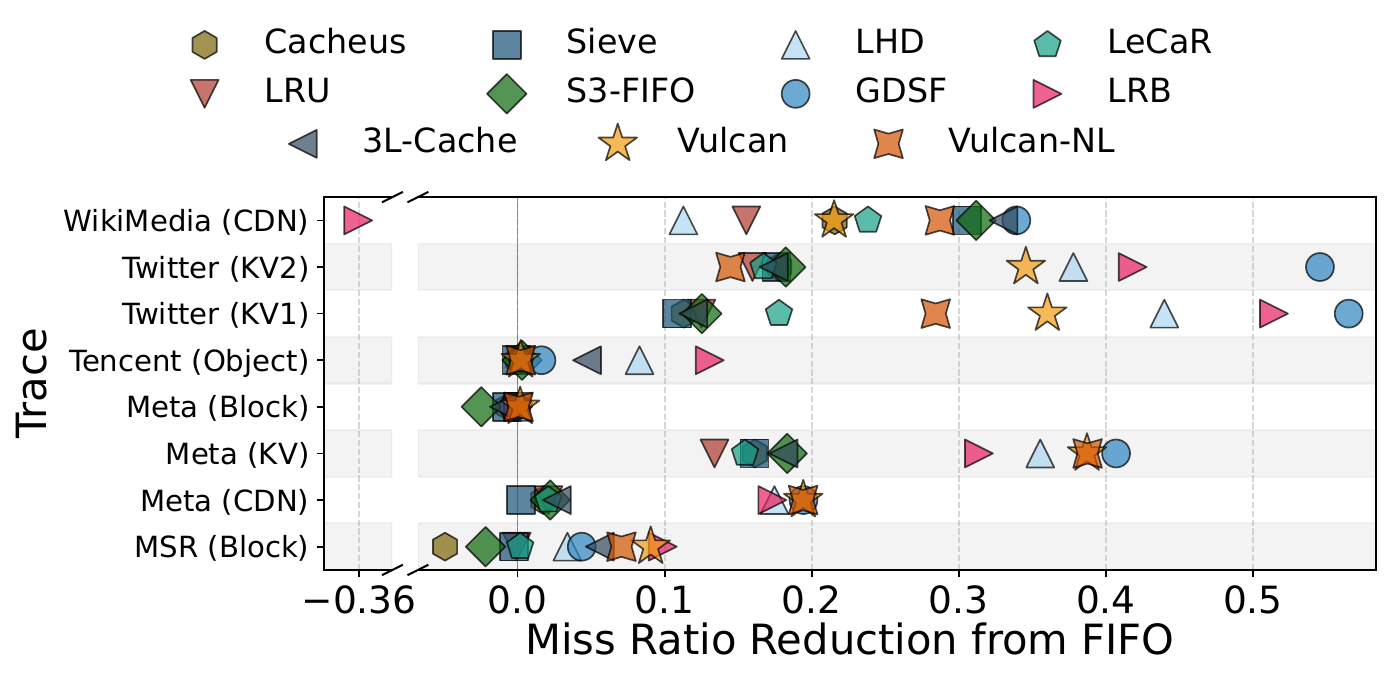}
        \caption{{Large cache size (10\%), objects of different sizes}}
        \label{fig:eviction-10pct}
    \end{subfigure}
    \hfill
    \begin{subfigure}[t]{0.49\textwidth}
        \centering
        \includegraphics[width=\linewidth]{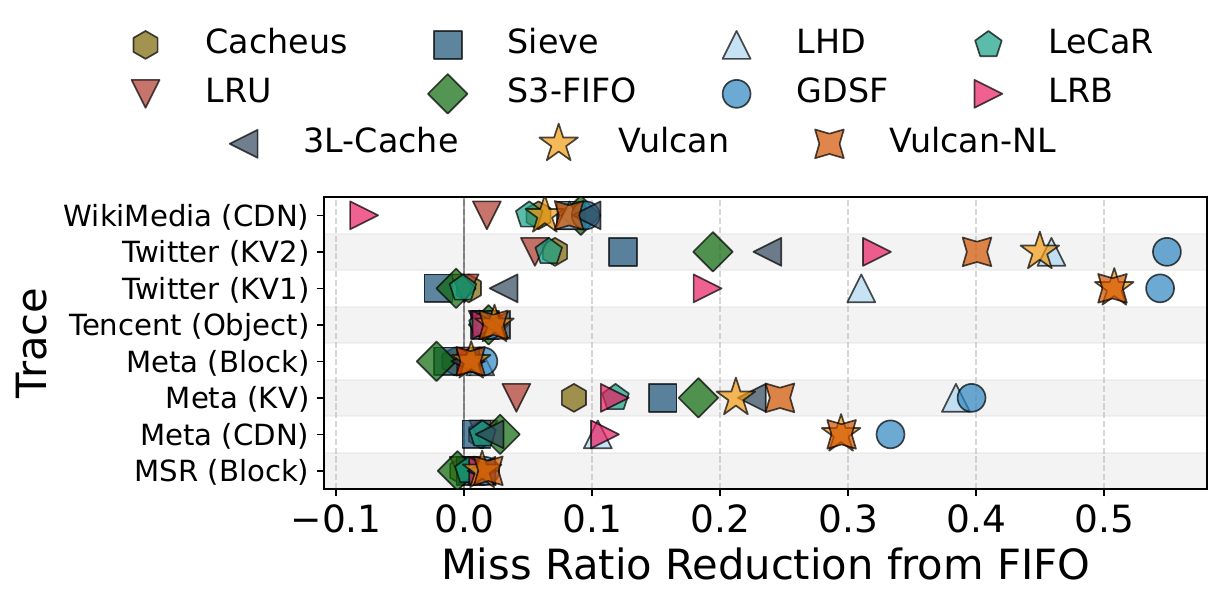}
        \caption{{Small cache size (0.1\%), objects of different sizes}}
        \label{fig:eviction-0.1pct}
    \end{subfigure}

    \caption{Performance of instance-specialized heuristics versus baselines for two classes of instances. Additional results appear in \autoref{app:caching-results}.}
    \Description{Two scatter plots compare miss-ratio reduction relative to FIFO for Vulcan, Vulcan without listeners, and nine baseline cache-eviction algorithms across eight workload traces; farther right indicates a larger reduction in miss ratio and therefore better performance. The top-to-bottom trace order is WikiMedia CDN, Twitter KV2, Twitter KV1, Tencent Object, Meta Block, Meta KV, Meta CDN, and MSR Block. The first plot uses a large cache equal to 10\% of the trace footprint, while the second uses a small cache equal to 0.1\%. For the large-cache setting, Vulcan is best on Meta Block, approximately matches the best baseline on Meta CDN, and is close to the best result on MSR Block. For the small-cache setting, Vulcan approximately matches the best baseline on Tencent Object and is close to the best result on MSR Block. Across traces, the identity of the best-performing algorithm varies substantially.}
    \label{fig:eviction-results}
\end{figure*}

Cache eviction is central to performance in CDNs~\cite{akamai-network,cachesack-atc22,robinhood-osdi18,caching-delayed-hits,lrb,fb-photo-cache} and in-memory stores~\cite{memcached,redis}.
As \autoref{subsec:bg:instance-optimality} showed, access patterns and object characteristics vary substantially across deployments, making instance-specialized eviction heuristics valuable.
Evicting an object from a pool of candidates maps naturally to a \Rank task (\autoref{sec:design:interfaces}): objects are scored by cache utility and the lowest-ranked object is evicted when space is needed.
We evaluate whether \system can synthesize such heuristics for individual cache deployments.

\underline{\textbf{Template}}.
We expose a \Rank scoring function over a priority queue of cached objects; per-object metadata (age, size, last access time, recent eviction history; see \autoref{tab:cache-features})\footnote{Note that the caching template shown in \autoref{lst:e2e-example} was simplified for exposition and contains only two raw signals.} is stored in a \libsystem feature store available to the function.
To reduce overhead, we use a lazy update strategy: scores are recomputed only when an object is accessed, not on every cache event.

\underline{\textbf{Evaluator}}.
We use libCacheSim~\cite{libcachesim} to simulate eight deployment-specific traces from Microsoft~\cite{msr-dataset}, Meta~\cite{cachelib}, Twitter~\cite{twitter-kv}, Tencent~\cite{tencent-dataset}, and Wikimedia~\cite{wikimedia}.
For each trace, we consider two cache sizes (10\% and 0.1\% of the trace footprint) and two object-size settings (uniform and variable), yielding 32 instances total.
During synthesis, candidates are scored on the first 1M requests (with a timeout of 100\,s to encourage efficient heuristics); the final heuristic is evaluated over the full trace, measuring miss-ratio reduction (MRR) over FIFO (\autoref{sec:bg}).

\underline{\textbf{Results}}.
{\emergencystretch=1em Our nine state-of-the-art eviction baselines are GDSF~\cite{gdsf}, S3-FIFO~\cite{s3-fifo}, SIEVE~\cite{sieve}, LHD~\cite{lhd}, Cacheus~\cite{cacheus}, LRB~\cite{lrb}, 3L-Cache~\cite{3l-cache}, LeCaR~\cite{lecar}, and LRU.
We use ShinkaEvolve~\cite{shinkaevolve} as the code-generation framework for generating heuristics for each of the 32 instances described above.
We also run an ablation where we disable listeners (named \system-NL).\par}

\noindent{\bf Overall results.}
Across the 32 instances, \system-generated heuristics outperform {\it all} baselines in {two instances and rank among the top three heuristics for 13 instances. Furthermore, \system heuristics are within 1\% of the best baseline on six instances and within 5\% on an additional six instances.}
A single implementation achieves all these results automatically after a one-time development effort.

\noindent{\bf Breakdown of results.}
\autoref{fig:eviction-10pct} and \autoref{fig:eviction-0.1pct} show miss-ratio improvements over FIFO for sixteen instances.
For the 10\% cache configuration, \system outperforms all baselines on the Meta block trace, matches the performance of the best (GDSF) on the Meta CDN trace, and {comes within 1\% of the best (LRB) on the MSR trace.} 
On the MSR block trace, the synthesized heuristic (\autoref{lst:cache-heuristics}) blends raw access count with an EWMA-smoothed count, using a ghost multiplier to emphasize recently evicted objects.
For the Meta block trace, it instead measures per-object access intensity and applies an additive penalty for ghost-list objects rather than a multiplicative one.
For the 0.1\% cache configuration, {\system matches the performance of the best baseline (3L-Cache) on the Tencent trace} and comes within 1\% of the best (GDSF) on the MSR block trace. 
\system achieves over 2$\times$ lower miss ratios than strong baselines such as SIEVE and S3-FIFO on specific instances: for instance, on the Twitter KV1 trace  (0.1\% cache size, variable object sizes) \system's miss ratio is 35.4\%, compared to 73.4\% for SIEVE and 72.5\% for S3-FIFO.
On the Tencent object trace, \system generates a GDSF-inspired heuristic similar to the MSR 10\% case, fine-tuning the constants for frequency and size to suit the Tencent workload and smaller cache.
{Detailed results for the remaining instances are in \autoref{app:caching:same-size-results} (\autoref{fig:eviction-10pct-nosize} and \autoref{fig:eviction-0.1pct-nosize})}.

\input{code-snippets/cache-heuristics}

\noindent{\bf Impact of listeners.}
Removing listeners only slightly degrades performance: \system-NL still outperforms all the baselines on {two instances} but trails the full \system by 0.84\% in average MRR. This difference is {small for this use case} since eviction heuristics are invoked extremely frequently, making complex listeners prohibitively expensive; computationally heavy heuristics are likelier to time out during evolution, {biasing the search towards simpler listeners}.

\noindent{\bf Cost of synthesis.} Across all 32 instances, synthesis cost \$272 for LLM calls ($\approx$ 36.5M input, 8.6M output tokens) and consumed 85 CPU-hours to evaluate candidate solutions.

    \noindent{\bf Runtime cost of \system heuristics.} \autoref{tab:slowdown} shows that \system policies' runtime overhead is comparable to that of 3L-Cache~\cite{3l-cache}, with which our policies are very competitive. Additionally, LRB~\cite{lrb}, which outperforms \system in six instances, takes 27$\times$ as long as \system on average. \system's cost can be reduced further by sampling: for instance, approximating the \system eviction policy for Meta KV with variable object sizes (10\% cache size) using $k=5$ samples reduces the cost from $15\times$ FIFO to $5\times$ FIFO while largely preserving miss rates.

\begin{table}[tb]
\centering
\caption{\textbf{Computational cost of eviction heuristics.} Average (and range) of execution time to run a set of traces (scenarios) in a single-threaded simulator. Higher = more expensive.}
\label{tab:slowdown}
\footnotesize
\begin{tabular*}{\linewidth}{@{\extracolsep{\fill}}lr@{}}
\toprule
\textbf{Algorithm} & \textbf{Execution time vs.\ FIFO} \\
\midrule
LRU                      & 1.0$\times$ (0.5--2.1$\times$) \\
SIEVE~\cite{sieve}        & 1.1$\times$ (0.6--2.0$\times$) \\
LeCaR~\cite{lecar}        & 1.8$\times$ (0.8--3.0$\times$) \\
S3-FIFO~\cite{s3-fifo}     & 2.1$\times$ (1.4--3.8$\times$) \\
GDSF~\cite{gdsf}          & 3.9$\times$ (2.5--7.1$\times$) \\
Cacheus~\cite{cacheus}    & 4.1$\times$ (2.3--7.9$\times$) \\
LHD~\cite{lhd}           & 4.1$\times$ (2.2--7.1$\times$) \\
\textbf{Vulcan}          & \textbf{13.4$\times$ (7.4--22.5$\times$)} \\
3L-Cache~\cite{3l-cache}  & 14.4$\times$ (5.3--38.2$\times$) \\
LRB~\cite{lrb}           & 361.2$\times$ (19.8--614.6$\times$) \\
\bottomrule
\end{tabular*}
\end{table}

%% file: code-snippets/cache-heuristics.tex
\begin{figure}[tb]
\begin{codebox}[basicstyle=\ttfamily\footnotesize,
    escapechar=|,
    caption={Simplified snippets of evolved heuristics for MSR and Meta block traces: actual heuristics are 29 and 50 lines of code.},
    label={lst:cache-heuristics}]
// MSR block trace
freq     = |\constcolor{0.7}|*count + |\constcolor{0.3}|*EWMA(count, alpha=|\constcolor{0.2}|)
size_pen = log2(size)
ghost_mult = |\constcolor{1}| + |\constcolor{2}|*ghost_count
utility  = (last_access/|\constcolor{1e6}| + |\constcolor{10}|*freq - size_pen) * ghost_mult

// Meta block trace
intensity  = count / (last_access - insert_time)
freq       = log2(count) * (|\constcolor{10}|*intensity)
size_fac   = |\constcolor{1}| / log2(size)
ghost_bonus = |\constcolor{1000}| * (|\constcolor{1}| + log2(ghost_count))
utility    = (|\constcolor{0.6}|*last_access + |\constcolor{400}|*freq) * size_fac + ghost_bonus
\end{codebox}
\end{figure}

%% file: sections/case-studies/memory-tiering.tex
\subsection{Memory Tiering}
\label{sec:tiering}
Memory tiering expands effective memory capacity by placing pages across fast local DRAM and slower tiers such as CXL-attached memory or NVM. The central policy problem is deciding \textit{which pages} should occupy scarce DRAM: ideally, frequently accessed pages should be promoted to DRAM, while colder pages remain in slower tiers. We use \system to synthesize page promotion heuristics and seek to answer whether they generalize from an emulated tiered-memory setup to real CXL hardware.

\underline{\textbf{Manually designed heuristics}}.
Existing memory tiering systems~\cite{hemem,tpp2023,memtis2023,arms} use sampled access information -- from mechanisms such as Intel PEBS~\cite{pebs2018} or page-table accessed-bit scanning~\cite{damon2019} -- to estimate page hotness and migrate pages accordingly. Memtis~\cite{memtis2023} represents one class of designs: it promotes pages whose sampled access count exceeds a dynamically adjusted hotness threshold and demotes pages once their cooled access counts fall below it. ARMS~\cite{arms} represents another class, using a ranking-based approach instead: it computes per-page scores from multi-timescale access averages, ranks pages by those scores, and migrates the highest-ranked pages to DRAM.

\underline{\textbf{Template}}.
We formulate the page migration problem as a \Rank task in \system, similar to ARMS~\cite{arms}: the goal is to estimate each page's hotness and rank pages accordingly. The raw signals provided for this task include per-page access counts over a time window (500\,ms) and the observed bandwidths of the hot and cold tiers. The developer-written template invokes \libsystem's \Rank every 1\,s.

\underline{\textbf{Evaluator}}. We score candidate heuristics by running a single workload on a two-tier memory \emph{emulator}. The emulator runs on a dual-socket Xeon CloudLab~\cite{cloudlab} \verb|c220g5| node: the near tier is socket 0's DRAM, and the far tier is socket 1's DRAM with its uncore frequency -- the clock governing the mesh interconnect, last-level cache, and memory controllers -- clamped to the minimum ratio.
This inflates remote-access latency to \textit{emulate} that of CXL-attached memory, while avoiding the cost of running heuristic search directly on scarce CXL hardware. After synthesis, we analyze whether the resulting heuristic generalizes to other workloads by testing it on our CXL-based testbed.

\noindent\textbf{Evolution setup}. We seed the search with a ``promote nothing'' heuristic that scores every page as 0 (\ie no migrations happen). We then run OpenEvolve~\cite{openevolve} for 100 iterations, where heuristic candidates are scored by their end-to-end goodput on a specific application; we then use the best heuristic in the generalization study.

\noindent\textbf{Generalization to real CXL hardware and unseen workloads.} We take this single heuristic synthesized against a single application on an emulator and, without further tuning, evaluate it on a real CXL machine across a diverse set of workloads. Our experimental setup pairs a dual-socket Xeon host with a Micron CZ120 CXL 2.0 memory expander (256\,GiB), connected to socket 0 over PCIe Gen5 x8 and exposed as a CPU-less NUMA node. The test workloads span three domains: GUPS~\cite{gups-benchmark} (memory microbenchmark used for training), GapBS Betweenness Centrality and PageRank~\cite{gap-benchmark-tiering} (graph analytics), and Silo TPCC~\cite{silo-memory-tiering} (in-memory database). All results are reported normalized to ARMS~\cite{arms}.

\begin{figure}[tb]
\begin{codebox}[basicstyle=\ttfamily\footnotesize,
    escapechar=|,
    caption={Memory tiering heuristic synthesized by \system.},
    label={lst:tiering-heuristics}]
fs.add_listeners(accesses, {EWMA({0.95, 0.35})});
double score_fn(|\systemnameplain|::store& fs, int64_t obj_id):
    double fast = fs.get_ewma(accesses, obj_id, 0.95);
    double slow = fs.get_ewma(accesses, obj_id, 0.35);
    double accel = (slow > 0.001) ? (fast / slow) : 1.0;
    return fast * (1.0 + 0.55 * min(accel, 2.5));
\end{codebox}
\end{figure}

\begin{figure}[tb]
    \centering
    \includegraphics[width=0.9\linewidth]{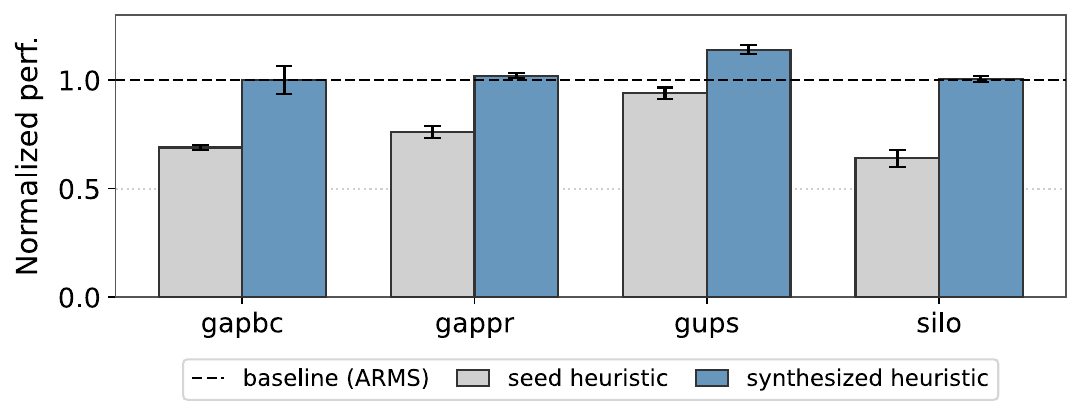}
    \caption{Performance of the \system-synthesized heuristic compared with the no-op seed heuristic.}
    \label{fig:generalization}
    \Description{Bar chart comparing the no-op seed heuristic and the Vulcan-synthesized heuristic across GapBC, GapPR, GUPS, and Silo, with performance normalized to ARMS; the dashed horizontal line at 1.0 represents ARMS performance. The no-op seed achieves approximately 0.68, 0.76, 0.95, and 0.65 times ARMS performance on GapBC, GapPR, GUPS, and Silo, respectively. The synthesized heuristic achieves approximately 1.00, 1.02, 1.14, and 1.01 times ARMS performance, respectively. Error bars show variability across runs.}
\end{figure}

\underline{\textbf{Results}}. \autoref{lst:tiering-heuristics} shows the synthesized heuristic, \system-GUPS, which used the GUPS benchmark~\cite{gups-benchmark} as the evaluator. \autoref{fig:generalization} shows its performance on the CXL testbed. From \autoref{lst:tiering-heuristics}, we observe that the heuristic uses the ratio of two EWMAs to derive a metric called the \textit{acceleration}, which it uses to make promotion decisions. We observe that this synthesized heuristic outperforms ARMS by 14.2\% on GUPS on CXL and matches the performance of ARMS on all other workloads, showing that heuristics synthesized on an emulator can generalize to real hardware.

\begin{figure}[tb]
    \centering    \includegraphics[width=0.95\linewidth]{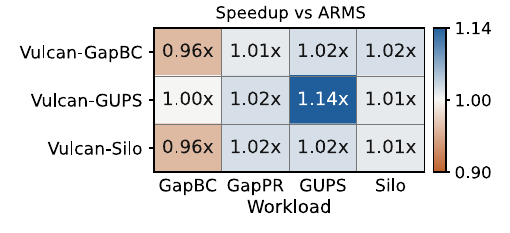}
    \caption{Generalization of \system-synthesized heuristics.}
    \Description{Heatmap showing how three Vulcan-synthesized heuristics generalize across four workloads, with each cell reporting performance normalized to ARMS. Columns are GapBC, GapPR, GUPS, and Silo. Vulcan-GapBC achieves 0.96, 1.01, 1.02, and 1.02 times ARMS performance, respectively. Vulcan-GUPS achieves 1.00, 1.02, 1.14, and 1.01 times ARMS performance. Vulcan-Silo achieves 0.96, 1.02, 1.02, and 1.01 times ARMS performance. Thus, the synthesized heuristics generally match or exceed ARMS, with the largest improvement, 1.14 times, occurring for Vulcan-GUPS on GUPS.}
    \label{fig:heatmap-tiering}
\end{figure}

In \autoref{fig:heatmap-tiering}, we repeat this experiment, using other workloads for training during evolution to obtain heuristics \system-GapBC and \system-Silo. We observe that all three heuristics meet or exceed the performance of ARMS for GapPR, GUPS, and Silo workloads. The only exception to this trend is \system-GapBC, which only achieves 0.96$\times$ the performance of ARMS on its training workload of GapBC; we hypothesize that this could be because GapBS-BC is a workload that performs breadth-first search over a graph, and hence can produce random pointer-chasing patterns. Such patterns may not be well captured by heuristics that score pages individually. Synthesizing these heuristics cost \$36 in LLM calls (5.8M input, 0.9M output tokens) and required a combined total of $7.8$ node-hours on our emulated CXL setup.

%% file: sections/08-discussion.tex
\section{Discussion}
\subsection{Developer Effort in \system}
\label{sec:discussion:system-integration}
\system changes not only the amount of developer effort required for synthesis but also \textit{when} that effort is incurred. With unconstrained synthesis, each synthesized heuristic is a new free-form implementation: before deployment, a developer must determine \textit{where} changes were introduced, which external systems the heuristic interacts with, and whether those interactions are correct. With \system, developers instead define this external-facing boundary before the search begins, {through the \textit{template}}.

Setting up the template requires developers to specify which raw signals are available to the policy, configure a \libsystem feature store to collect and update these signals, and recast the policy's decision logic using \Rank/\Value {(\autoref{sec:design})}. {\autoref{tab:developer-effort} quantifies this setup effort for the use cases we study.}

\begin{table}[tb]
\footnotesize
\centering
\caption{\textbf{Developer effort to set up templates.}}
\label{tab:developer-effort}
\begin{tabular*}{\linewidth}{@{\extracolsep{\fill}}lcccc@{}}
\toprule
\multirow{2}{*}{\textbf{Metric}} & \textbf{Spot VM} & \textbf{Spot VM} & \textbf{Cache} & \textbf{Memory} \\
& \textbf{(single)} & \textbf{(multi)} & \textbf{eviction} & \textbf{tiering} \\
\midrule
Raw signals    & 8      & 11           & 6     & 3     \\
{Template} (LoC) & 62     & 104          & 74    & 72    \\
\bottomrule
\end{tabular*}
\end{table}

\autoref{tab:developer-effort} shows that this refactoring is a modest one-time cost: our case studies required exposing only 3--11 raw signals and writing 62--104 lines of {task-specific code for the template}. This effort is then amortized across instances requiring specialization. For example, in cache eviction (\autoref{sec:caching}), the same 74-line template was reused to synthesize heuristics for all 32 instances.

This fixed synthesis boundary also makes synthesized artifacts more predictable: every heuristic implements the predefined \Rank or \Value interface and operates over the same declared signals. Thus, even when a developer chooses to inspect a synthesized heuristic in \system, they already know the shape of what they are looking at, substantially reducing the effort required to review each synthesized heuristic.

\subsection{\system's \Rank and \Value Interfaces}
\label{sec:discussion:interfaces}
\subsubsection{Scope and Limitations.}
While our interfaces capture the core decision logic of a broad class of systems policies (\autoref{tab:types-of-policies}), we do not claim they are universal. More complex problems can often be decomposed into \Rank and \Value subproblems: as we demonstrate in our multi-region spot-VM case study (\autoref{sec:spot-vm-scheduling-multi}), a \Value function decides whether to migrate, and a \Rank function then selects the destination region only if the first policy \textit{chooses} to migrate. Such decompositions, however, impose explicit structure on how sub-policies interact: in this setting, the first policy cannot decide to migrate \emph{because} an attractive destination exists. As the number of coupled decisions grows, decomposing them into separate components requires increasingly elaborate coordination between them. For recurring patterns of this kind, extending \system with new interfaces may provide cleaner abstractions than further composition.

\subsubsection{Extending \system with New Interfaces.}
The key requirement for a new interface is that its synthesized decision logic remain stateless. This preserves \dsl's deliberately restrictive design: synthesized programs cannot allocate heap memory or maintain persistent mutable state, enabling strong execution-safety guarantees by construction. For example, a new \texttt{COMPARE} interface could choose between two alternatives based on their features jointly, unlike \Rank, which scores each candidate independently. Similarly, a \texttt{CLASSIFY} interface could return an element from a finite set of actions directly, rather than encoding categorical choices through \Value's scalar output. More generally, recurring decision abstractions that admit stateless decision logic could be incorporated into \system as additional interfaces.

%% file: sections/09-related-work.tex
\section{Related Work}
\label{sec:related}
\noindent{\bf Traditional learning-based policies.}
Prior ML-based systems policies, particularly neural policies~\cite{orca,c3,heimdall,lrb,glcache,3l-cache}, have demonstrated that data-driven approaches can outperform fixed heuristics on specific instances.
However, neural approaches introduce significant practical challenges: opaque behavior that complicates debugging~\cite{metis-sigcomm20}, complex training and deployment pipelines~\cite{lake,kml-lib}, inference overheads in the control path~\cite{lake,liteflow-sigcomm22}, and safety concerns that limit adoption~\cite{guardrails-hotos}.
\system takes a different stance: it avoids neural inference in the hot path entirely, confining learning to an offline search over small, interpretable LLM-generated code snippets.

\noindent{\bf AI-driven algorithm and heuristic design.}
Recent work has explored using LLMs to discover new heuristics via evolutionary search~\cite{funsearch,alphaevolve}.
One line of work modifies existing human-designed algorithms incrementally, such as automatic BBR variants~\cite{cc-tweak-bbr,barbarians}.
A complementary line focuses on search strategies: Glia~\cite{glia} employs multiple specialized LLM agents, and Robusta~\cite{robusta} uses counterexamples to harden candidate heuristics.
As we show in \autoref{sec:why-llms-fail}, unconstrained approaches of this kind can produce safety and integration bugs.
These search strategies are complementary to \system: any of them, combined with \system's \Rank and \Value interfaces and listeners, can yield similar benefits.

\noindent{\bf Domain-specific languages for execution safety.}
Restricted languages such as Rust and eBPF~\cite{ebpf, gbadamosi-ebpf} require complex verifiers and carry steep learning curves; even then, they do not enforce leak-freedom~\cite{rust-book-memory-leak, gbadamosi-ebpf}, which is too restrictive a guarantee for general programs.
General-purpose verifiers~\cite{dafny, verus} can encode execution-safety properties as pre- and post-conditions, but require substantial per-program annotation effort, including loop invariants, termination arguments, and framing conditions~\cite{ironfleet, hyperkernel}.
\dsl takes the opposite extreme: by restricting LLM-written code to simple stateless functions implementing \Rank or \Value, it ensures P1--P3 by construction without any annotations.

\section{Conclusion}
This paper presents \system, a framework for synthesizing systems resource-management heuristics that are both performant and safe to deploy. Across three use cases, \system discovers heuristics that are competitive with strong human-designed baselines, while avoiding the execution-safety and system-integration problems that arise with unrestricted code generation.

More broadly, \system's results suggest that safe synthesis need not come at the cost of meaningful policy exploration. An important next step is to explore how far this boundary can be expanded, for example by using stronger end-to-end semantic verification to safely expose a larger synthesis surface. Another direction is to formalize what constitutes an \emph{instance} in practice: when to synthesize a new heuristic rather than reuse an existing one, what should trigger re-synthesis, and how to handle transitions between policies.

%% file: utils/ack.tex
\section{Acknowledgements}
This material is based upon work supported by the U.S. National Science Foundation (NSF) under Grant Number 2326576. This research was also supported by the \href{https://infraai.center/}{InfraAI Center} at The University of Texas at Austin. Dwivedula was supported by an Amazon AI Fellowship. We are also grateful to our shepherd for their thoughtful feedback, which helped improve the clarity and presentation of this paper.

%% file: appendix/appendices.tex
\appendix
\input{appendix/appendixA-llms-fail}
\input{appendix/appendixB-cache-rank}
\input{appendix/appendixC-artifact}

%% file: appendix/appendixA-llms-fail.tex
\section{Challenges with LLM-Based Heuristic~Generation}
\label{app:why-llms-fail}
To evaluate whether current LLMs can synthesize effective systems heuristics that meet these requirements (\autoref{tab:safety-properties}), we used a state-of-the-art model, Claude Opus 4.5, to generate caching heuristics in libCacheSim~\cite{libcachesim}.

libCacheSim~\cite{libcachesim} is an event-driven cache simulator, in which new heuristics are defined by implementing event handlers for cache operations such as accesses, evictions, and insertions -- each of which has a specific function signature detailed in the documentation (a systems integration requirement).
Additionally, the libCacheSim engine also provides access to commonly used features -- such as total object counts, the current timestamp, and the number of hits/misses -- via raw pointers, which heuristics are allowed to read but must not mutate (another systems integration requirement).

\begin{figure}[tb]
    \centering
    \includegraphics[width=0.8\linewidth]{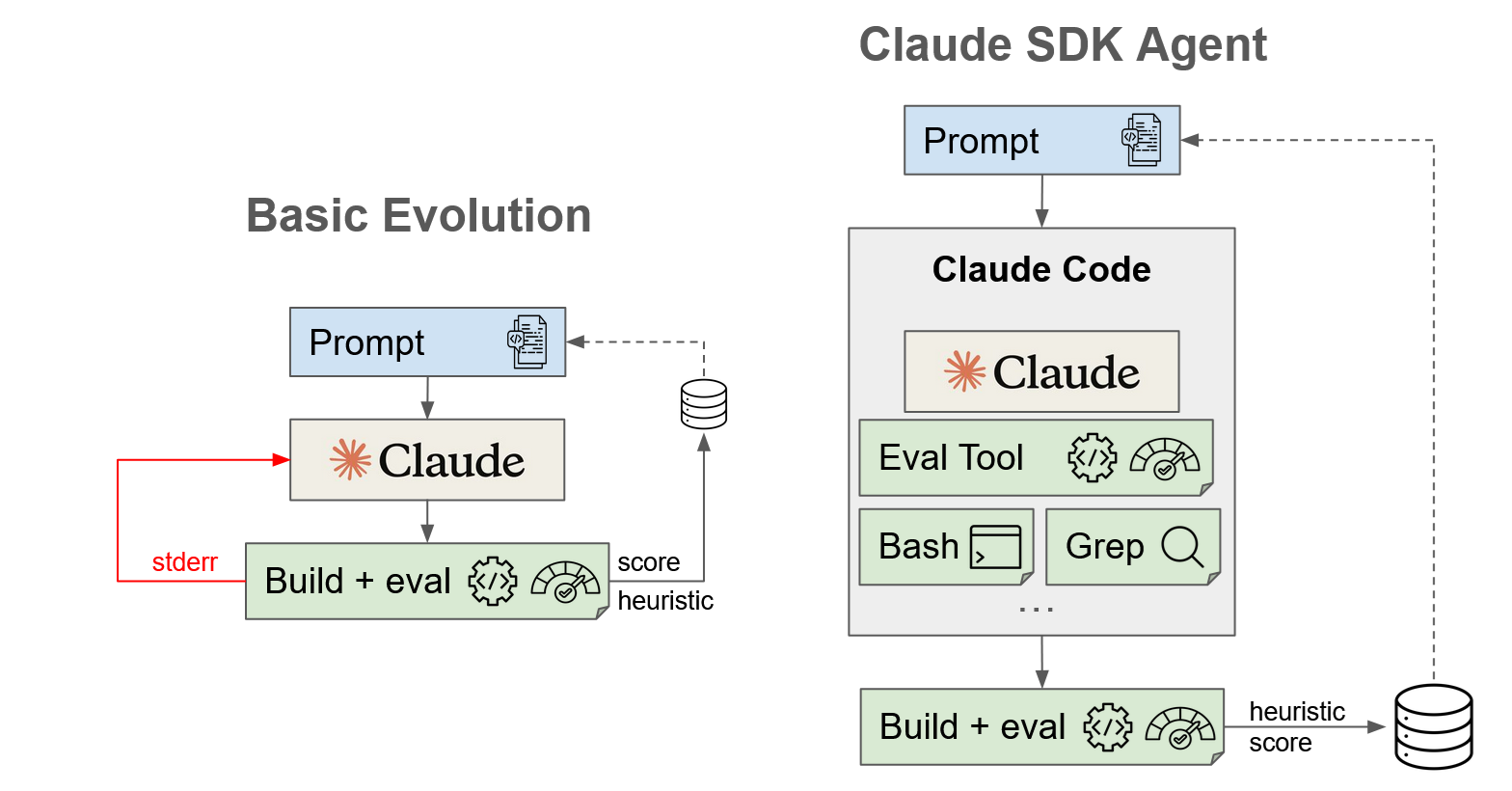 }
    \caption{Evolutionary search with a simple agent vs. Claude Code. The dotted line represents the start of a new iteration; every iteration is seeded with samples of the best-performing programs so far to help guide the search.}
    \Description{Side-by-side workflow diagrams comparing basic evolutionary search with a Claude Code agent. In basic evolution, a prompt is sent to Claude to generate a heuristic, which is built and evaluated; build or runtime errors are returned to Claude for correction, while successful heuristic scores are stored and used to seed later iterations. In the Claude Code workflow, the prompt is given to a Claude Code agent that can repeatedly inspect and modify the codebase and invoke tools including evaluation, Bash, and grep before producing a heuristic. The resulting heuristic is then built and evaluated, and its score is stored for use in subsequent iterations. Dotted boundaries indicate individual evolutionary iterations.}
    \label{fig:agent-comparison}
\end{figure}

\subsection{Attempt 1: A Simple Evolutionary Loop}
We implement an evolutionary loop using a simple LLM agent (inspired by prior work~\cite{funsearch,policysmith}), as shown in the left half of \autoref{fig:agent-comparison}. The agent is provided with a prompt describing libCacheSim and its interfaces, which it uses to generate a candidate heuristic.
The generated heuristic is then built and evaluated on a suite of ten traces from the CloudPhysics dataset~\cite{cloud-physics-dataset}. If a build or runtime error occurs, the agent receives the \verb|stderr| output to iteratively refine and correct its code. Once the heuristic compiles and executes successfully, both the implementation and its performance metrics (\ie miss rates) are stored in a database.
Each such iteration constitutes the \textit{generation} of a single heuristic. This process is repeated for fifty iterations, with higher-performing heuristics sampled from the database and incorporated into subsequent prompts to guide the search toward better designs.

Manual inspection revealed a recurring issue across all fifty heuristics. In libCacheSim, heuristics must report their per-object metadata footprint (in bytes) via a system variable used for memory accounting. Despite introducing additional per-object state, the LLM consistently under-reported this value. As a result, these heuristics exceeded their memory budget, yielding unfair comparisons against baselines that correctly account for memory usage and violating interface requirements (a system-integration bug).

Despite these issues, after fifty iterations, the evolutionary process produced heuristics that ``achieved'' a 3\%--5.1\% improvement over FIFO according to the evaluator.
However, these gains were partly driven by exploiting gaps in the evaluation harness (\ie under-reporting metadata usage).
Moreover, heuristics with memory safety bugs also passed without crashing, as the harness exercised only limited execution paths.
This shows that heuristics which compile, run successfully, and perform well on the evaluator may still violate the desired properties in \autoref{tab:safety-properties}.

\subsection{Attempt 2: Agentic Code Synthesis}
After initial experiments with simple agents showed that LLM-synthesized code often violates desired properties, we asked whether these issues persist in more capable agentic frameworks that augment LLMs with external tools.
To investigate, we repeated the heuristic synthesis experiment using Claude Code -- a widely used coding agent that equips an LLM with capabilities such as file manipulation, shell execution, and search.
Such agents may better avoid interface-compliance bugs observed earlier, as they can spawn sub-agents to reason about and understand the internals of the systems they interact with more effectively.

To test this hypothesis, we repeated the experiment using the Claude Code Agent SDK (Claude Opus 4.5) to synthesize each candidate heuristic, as shown in the right half of \autoref{fig:agent-comparison}. In addition to default tools (bash, search, read, etc.), we provided a custom \verb|evaluate| tool that builds the heuristic, runs it on CloudPhysics traces, and returns a JSON report of its performance relative to FIFO. Although the agent could execute these steps via bash, exposing them as a dedicated tool simplifies its workflow.

Using this more capable agentic setup, we ran the evolutionary process for ten iterations, producing ten heuristics. As in Attempt~1, each iteration generates a candidate via a loop of proposing, evaluating, and refining; however, unlike the earlier setup -- where the loop was limited to handling compiler and runtime errors -- each iteration here is a richer multi-turn interaction. The agent can explore the codebase, reason about interfaces, modify implementations, and invoke tools such as \verb|evaluate| multiple times before returning a candidate heuristic.

Generating each candidate heuristic with Claude Code took 11--28 minutes, consumed $\sim$2M tokens ($\sim$\$3.20 via AWS Bedrock), and involved an average of 57.4 tool calls. The resulting heuristics achieved a 36.4\%--46.8\% improvement over FIFO, as scored by the evaluator, significantly outperforming those found by the simpler agent.

They also exhibited diverse strategies: the best-performing heuristic uses a prioritization function combining frequency tiers, size-cost normalization, and recency protection, while the second-best applies a concave transformation (a sqrt--log blend) to weight access~frequency.

Unlike the previous setup -- where even modest gains were partly driven by reward hacking (\ie under-reporting metadata to bypass memory accounting) -- the heuristics produced by this agentic framework did not exploit that loophole. However, several still violated interface assumptions: for instance, several heuristics -- including the best-performing one -- overwrote \verb|next_access_vtime| (per-object state maintained by libCacheSim) to store the current time, violating the system interface by treating a read-only field as writable (a systems integration bug).
Additionally, a double-free bug (an execution safety bug) in the second-best heuristic went undetected by the evaluator; libCacheSim exposes two removal paths -- \verb|evict()|, which selects and removes objects, and \verb|remove()|, which deletes objects by ID. Since our evaluator exercised only the \verb|evict()| path, the bug in \verb|remove| went undetected.

\subsection{Summary}
This case study reinforces a growing body of evidence that LLMs routinely introduce bugs and safety violations when generating complex code~\cite{bigcodebench, livecodebench, security-bugs-study-2026, llm-bugs-code-translation} -- and that more capable agents are no exception, frequently requiring humans in the loop to produce reliable output~\cite{code-researcher,swe-polybench}.

While LLMs may be more successful at synthesizing code for self-contained coding tasks~\cite{human-eval, mbpp-benchmark, apps-benchmark}, systems heuristics are usually more complex: they may maintain internal state in complex data structures~\cite{maple-trees}, involve intricate state-transition logic~\cite{s3-fifo, cubic, arc}, or be dispersed across multiple files and control paths~\cite{cbmm}.
Allowing an LLM to synthesize or modify such heuristics end-to-end -- reasoning simultaneously about policy logic, state management, and mechanisms -- widens the attack surface for subtle bugs and property violations in the synthesized code.

%% file: appendix/appendixB-cache-rank.tex
\section{\Rank-based Cache Eviction Heuristics}
\label{app:caching-results}
\begin{figure*}[t]
    \centering
    \begin{subfigure}[t]{0.49\textwidth}
        \centering
        \includegraphics[width=\linewidth]{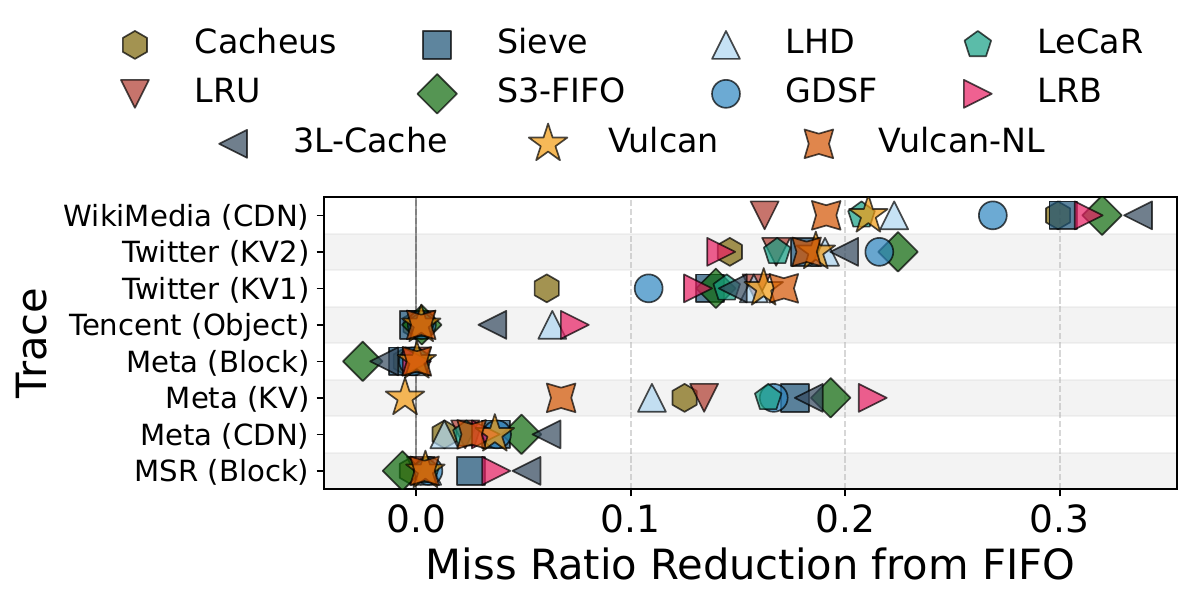}
        \caption{Large cache size (10\%), objects of the same size}
        \label{fig:eviction-10pct-nosize}
    \end{subfigure}
    \hfill
    \begin{subfigure}[t]{0.49\textwidth}
        \centering
        \includegraphics[width=\linewidth]{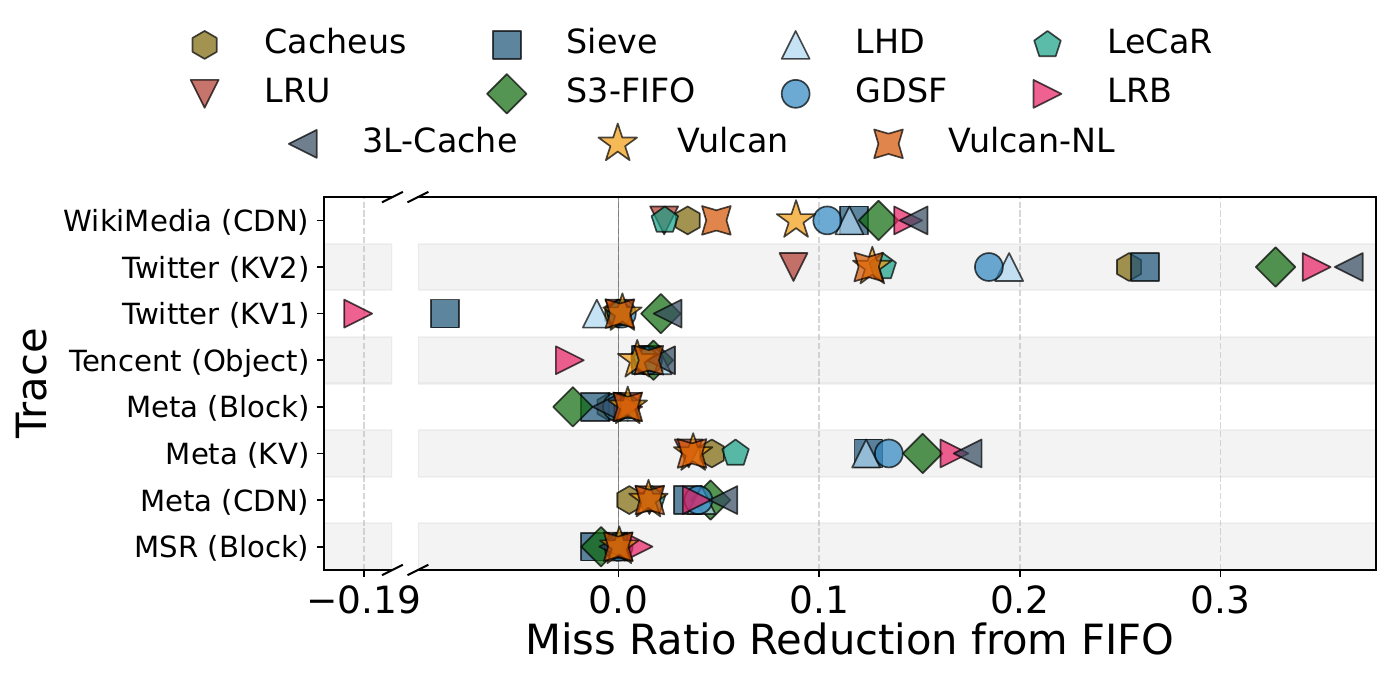}
        \caption{Small cache size (0.1\%), objects of the same size}
        \label{fig:eviction-0.1pct-nosize}
    \end{subfigure}

    \caption{Performance of instance-specialized heuristics versus baselines for two classes of instances.}
    \Description{Two scatter plots compare miss-ratio reduction relative to FIFO for Vulcan, Vulcan without listeners, and nine baseline cache-eviction algorithms across eight workload traces where all cached objects have the same size; farther right indicates a larger reduction in miss ratio and therefore better performance. The first plot uses a large cache equal to 10\% of the trace footprint, while the second uses a small cache equal to 0.1\%. For the large-cache setting, improvements vary substantially by trace: several methods achieve reductions above 0.2 on WikiMedia and Twitter workloads, while gains are much smaller on Meta CDN and MSR Block. For the small-cache setting, most methods cluster near zero on several traces, but larger improvements appear on Twitter KV2, Meta KV, and WikiMedia CDN. As in the different-object-size experiments, the best-performing algorithm varies across traces and cache sizes.}
    \label{fig:eviction-results-nosize}
\end{figure*}

\subsection{Results for Same-Size Instances}
\label{app:caching:same-size-results}

\autoref{fig:eviction-results-nosize} reports results for two classes of instances in which cached objects have uniform sizes.
As presented in \autoref{sec:caching}, we compare these heuristics with nine human-designed baselines and find that \system-generated heuristics are generally competitive against strong manually designed baselines.

\subsection{Synthesis Prompt}

The natural-language prompt used for the cache eviction case study (\autoref{sec:caching}) is shown in \autoref{lst:cache-prompt}. The prompt also includes dynamically selected examples of high-performing scoring functions from previous rounds, selected automatically by the evolution frameworks~\cite{openevolve,shinkaevolve}.

\begin{tcolorbox}[
  code={\refstepcounter{lstlisting}\label{lst:cache-prompt}},
  title=\textbf{Listing \thelstlisting: Prompt for synthesizing cache eviction~heuristics},
  breakable,
  colback=cyan!3,
  colbacktitle=cyan!10,
  colframe=cyan!50!black,
  boxrule=0.8pt,
  arc=2pt,
  left=6pt,
  right=6pt,
  top=6pt,
  bottom=6pt,
  coltitle=black,
  fonttitle=\bfseries,
  fontupper=\small,
]
\textbf{Context.}
You are a systems researcher designing a new cache eviction policy. Whenever the cache is full and a new object must be inserted, this policy is invoked to select which cached object to evict. The policy has access to per-object features such as object size, insertion time, last access time, and access count, along with global information about recently evicted~objects.

\medskip

\textbf{Policy Model.}
The policy is implemented using \texttt{\systemnameplain}, a framework that separates \emph{policy} from \emph{mechanism}. Your task is to implement a \emph{scoring function} that returns the \textbf{utility} of an object:
\[
\text{Higher utility} \Rightarrow \text{more valuable to keep}
\]
When eviction is required, select the object with the \emph{\mbox{lowest utility}}.

\medskip

\textbf{Lazy Priority Queue Execution Model.}
The scoring function is evaluated lazily:
\begin{itemize}
    \item A score is recomputed only when an object is inserted or accessed.
    \item The resulting score is cached in the priority queue.
    \item Idle objects are \emph{not} rescored over time.
\end{itemize}

Therefore, expressions such as $\texttt{curr\_time - last\_access}$ are ineffective because, at rescore time, $\texttt{curr\_time} \approx \texttt{last\_access}$.

Instead, use:
\begin{itemize}
    \item \texttt{last\_access} directly for recency,
    \item \texttt{insertion\_time} directly for age ordering,
    \item or other stable per-object signals.
\end{itemize}

\medskip

\textbf{Global Features.}
\begin{itemize}
    \item \texttt{f\_curr\_time (i64)}: Current logical time.
    \item \texttt{f\_ghost (i64)}: IDs of recently evicted~objects.
\end{itemize}

\medskip

\textbf{Per-Object Features.}
\begin{itemize}
    \item \texttt{f\_size (i64)}: Object size in bytes.
    \item \texttt{f\_insertion\_time (i64)}: Insertion timestamp.
    \item \texttt{f\_last\_access (i64)}: Most recent access timestamp.
    \item \texttt{f\_count (i64)}: Access count since insertion.
\end{itemize}

\medskip

\textbf{Listeners.}
Listeners determine how feature values are stored and queried. Examples include:
\begin{itemize}
    \item \texttt{RollingWindow(N)}
    \item \texttt{EWMA(\{alpha\})}
    \item \texttt{RollingPercentile(N)}
    \item \texttt{RollingCount(N)}
\end{itemize}

Example listener configuration:
\begin{lstlisting}[basicstyle=\ttfamily\footnotesize,escapeinside={(*}{*)}]
fs.add_listeners(
  f_last_access,
  {(*\systemnameplain*)::listeners::object::RollingWindow(5)}
);

fs.add_listeners(
  f_curr_time,
  {(*\systemnameplain*)::listeners::global::RollingWindow(1)}
);
\end{lstlisting}

\medskip

\textbf{Expected Output Format.}
The generated file must contain:
\begin{enumerate}
    \item Listener configuration
    \item Scoring function
\end{enumerate}

\medskip

\textbf{Example Scoring Function.}

\begin{lstlisting}[basicstyle=\ttfamily\footnotesize,escapeinside={(*}{*)}]
auto scoring_fn =
  [&](
    (*\systemnameplain*)::store& fs,
    int64_t obj_id
  ) -> double {
    double last_acc = fs.get_latest(
      f_last_access, obj_id
    );
    double count = fs.get_latest(
      f_count, obj_id
    );
    double ghost_count = fs.get_count(
      f_ghost, obj_id
    );
    double ghost_penalty = ghost_count * 1e6;
    return last_acc+(count*10)-ghost_penalty;
};
\end{lstlisting}
\end{tcolorbox}

%% file: appendix/appendixC-artifact.tex
\section{Artifact Appendix}
Our artifact contains: (a) the final discovered heuristics for each of our use cases, (b) orchestrator and plotting scripts to recreate the main results of the paper, (c) scripts to run heuristic discovery loops for each use case with off-the-shelf evolution frameworks~\cite{shinkaevolve,openevolve}, and (d) implementations of \libsystem and \dsl and documentation on how to extend and use these libraries to discover heuristics for new tasks.

\noindent Our artifact is available at the following links:
\begin{itemize}
    \item Main repository containing code for all four use cases. \begin{itemize}
        \item  \href{https://github.com/ldos-project/vulcan-usecases}{github.com/ldos-project/vulcan-usecases}
    \end{itemize}
    \item Common libraries: \begin{itemize}
        \item \href{https://github.com/ldos-project/libvulcan}{libVulcan} (DOI: \href{https://doi.org/10.5281/zenodo.22905536}{10.5281/zenodo.22905536})
        \item \href{https://github.com/ldos-project/anvil}{Anvil} (DOI: \href{https://doi.org/10.5281/zenodo.22905483}{10.5281/zenodo.22905483})
    \end{itemize}
    \item Discovered heuristics and precomputed results (DOI: \href{https://doi.org/10.5281/zenodo.20361337}{10.5281/zenodo.20361337})
\end{itemize}

Each use case has a separate repository branch. Check out the appropriate branch of \verb|vulcan-usecases| and follow its \verb|README.md|. For each use case, we provide precomputed results (e.g., JSON files and dumps) that were used to create the plots -- use these dumps directly or rerun the evaluation.

\subsection{Reproducing Spot VM Single-Region Results}
\begin{itemize}
    \item Branch: \verb|cbl-artifact|;
    \item \autoref{sec:spot-vm-scheduling-single};
    \item Plot(s) reproduced: \autoref{fig:spot-single-region-results}.
    \item Plotting script: \verb|python3 plot_main_result.py|.
    \item DOI: \href{https://doi.org/10.5281/zenodo.22905385}{10.5281/zenodo.22905385}
\end{itemize}

\subsection{Reproducing Spot VM (Multi-Region) Results}
\begin{itemize}
    \item Branch: \verb|cbl-multi-artifact|;
    \item \autoref{sec:spot-vm-scheduling-multi};
    \item Plot(s) reproduced: \autoref{fig:spot-multi-region-results}.
    \item Plotting script: \verb|plot_results.py|
    \item DOI: \href{https://doi.org/10.5281/zenodo.22905389}{10.5281/zenodo.22905389}
\end{itemize}

\subsection{Reproducing Caching Results}
\begin{itemize}
    \item Branch: \verb|caching-artifact|;
    \item \autoref{sec:caching};
    \item Plot(s) reproduced: \autoref{fig:eviction-results} and~\autoref{fig:eviction-results-nosize}.
    \item Plotting script: \verb|plot_workload_instances.py|
    \item DOI: \href{https://doi.org/10.5281/zenodo.22905373}{10.5281/zenodo.22905373}
\end{itemize}

\subsection{Reproducing Memory Tiering Results}
\textit{Note: for this use case, the artifact reproduces results from the emulated setup and requires CloudLab.}
\begin{itemize}
    \item Branch: \verb|tiering-artifact|;
    \item \autoref{sec:tiering};
    \item Plot(s) reproduced: \autoref{fig:generalization}.
    \item Plotting script: \verb|plot_improvement.py|
    \item DOI: \href{https://doi.org/10.5281/zenodo.22905376}{10.5281/zenodo.22905376}
\end{itemize}

\subsection{Running Evolution and Adding Use Cases}
Beyond reproducing the final results, each use case also includes the synthesis harness and prompts, allowing users to rerun the full discovery loop with an off-the-shelf evolution framework (OpenEvolve or ShinkaEvolve). Additionally, users who wish to use our framework for their use cases can find documentation in \href{https://github.com/ldos-project/libvulcan}{libvulcan}.